\documentclass[a4paper,12pt]{article}

\usepackage{amsmath}
\usepackage{amssymb}
\usepackage{graphicx}
\usepackage{cite}

\usepackage[margin=1in]{geometry}

\newcommand{\ice}[1]{\relax}

\newcounter{sectio}

\newcommand{\ocite}[1]{\cite{#1}}

\newcommand{\re}[1]{(\ref{#1})}

\newcommand{\nnb}{\nonumber}

\newcommand{\foot}[1]{\footnote{#1}}

\newcommand{\chap}[1]{\section{#1}}
\newcommand{\sect}[1]{\subsection{#1}}

\ice{
\d was aliased to  \delta, now i replace it with \delta  everywhere
using a command:

}

\newcommand{\beq}{\begin{equation}}
\newcommand{\eeq}{\end{equation}}

\newcommand{\bea}{\begin{eqnarray}}
\newcommand{\eea}{\end{eqnarray}}

\newcommand{\EQN}{\label}

\newcommand{\ba}{\begin{array}}
\newcommand{\ea}{\end{array}}

\newcommand{\Wu}{W_{\!{}_U}}
\newcommand{\Fu}{F_{\!{}_U}}
\newcommand{\dsp}{\displaystyle}
\newcommand{\eol}{\endgraf\noindent}
\newcommand{\llp}{{}'\!\!\!\!}
\newcommand{\lp}{{}'}
\newcommand{\aseq}{\mathop{=\!\!=\!\!=}}

\newcommand{\as}{\aseq_{\rho\to0}}
\newcommand{\asq}%
{\,\aseq_{{\!\!}
Q\to\infty}\,}

\newcommand{\T}{{\cal {T}}}

\newcommand{\li}{\widetilde <}
\newcommand{\gi}{\widetilde >}
\newcommand{\Wt}{\widetilde W}

\newcommand{\Dt}{\widetilde \D }
\newcommand{\ct}{\widetilde c}
\newcommand{\Ct}{\widetilde C}
\newcommand{\Rt}{\widetilde R}
\newcommand{\up}{\!\uparrow}
\newcommand{\g}{\gamma}
\newcommand{\G}{\Gamma}
\newcommand{\Ru}{R_{{}_U}}

\newcommand{\Rm}{R^{-1}}
\newcommand{\D}{\Delta}

\newcommand{\Dm}{\Delta^{-1}}
\newcommand{\Du}{\Delta_{{}_U}}
\newcommand{\Di}{\Delta_{{}_I}}

\newcommand{\ep}{\epsilon}
\newcommand{\de}{\delta}

\newcommand{\LG}{{\cal{L}}_{\Gamma}}
\newcommand{\VG}{{\cal{V}}_{\Gamma}}
\newcommand{\EG}{{\cal{E}}_{\Gamma}}
\newcommand{\Lg}{{\cal{L}}_{\gamma}}
\newcommand{\Vg}{{\cal{V}}_{\gamma}}

\newcommand{\gs}{\widetilde>}
\newcommand{\ls}{\widetilde<}
\newcommand{\tld}{\sim}
\newcommand{\ed}{\end{document}}

\renewcommand{\b}{\backslash}

\numberwithin{equation}{section}

\title{\bf \large  
 COMBINATORICS OF $\mathbf{R}$-, $\mathbf{R^{-1}}$-, AND $\mathbf{R^*}$-OPERATIONS
\\
AND 
\\
ASYMPTOTIC EXPANSIONS OF FEYNMAN INTEGRALS 
\\
IN THE LIMIT OF LARGE  MOMENTA AND MASSES\thanks{
This is  an Archive  copy of  a preprint MPI-Ph/PTh 13/91 issued 
by the Max-Plank-Institute  f\"ur Physik (Munich, Germany)  in March of 1991.
See the Comments Section in the very end. 
}}
\author{K.G. Chetyrkin  \\
Institut f\"ur Theoretische Teilchenphysik, Karlsruhe
\\
  Institute of Technology (KIT), Germany
	}

 \date{ }
\begin{document}
\maketitle

\begin{abstract}
  A generalization of the forest technique procedure --- the
  $R^{-1}$-operation---is elaborated and then employed to treat a variety of
  problems. First, it is employed to reveal the underlying simple structure of
  the Bogoliubov-Parasiuk renormalization prescription based on momentum
  subtractions. Second, we use this structure to derive a generalized
  Zimmermann identity connecting two different renormalized versions of a
  given Feynman integral. Third, the recursive procedure to minimally subtract
  the ultraviolet and infrared divergences from euclidean, dimensionally
  regularized Feynman integrals---the $R^*$-operation--- is simplified by
  reformulating it in terms of the R-operation alone.  The new formulation is
  shown to lead immediately to a simple and regular algorithm for evaluating
  the overall ultraviolet divergences of arbitrary dimensionally regularized
  Feynman integrals, (including the ones appearing in two-dimensional
  field-theoretical models), the algorithm neatly reducing the problem to
  computing some massless propagator-type integrals. Finally, we construct a
  brief and concise proof of a general theorem which gives an explicitly
  finite large momenta and/or masses asymptotic expansion of an arbitrary
  (minimally subtracted) euclidean Feynman integral.

\end{abstract}

\newpage

\tableofcontents 

\newpage 

\setcounter{sectio}{1}
\setcounter{thm}{0}
\setcounter{lem}{0}
\setcounter{equation}{0}

\chap{INTRODUCTION}
One of  cornerstones of the local  quantum  field theory is
renormalization, i.e.   proper identification and
self-consistent subtraction of infinities that plague
perturbation series. A  rigorous all-order treatment of the
renormalization problem was
began with the classic papers of  Bogoliubov  and
Parasiuk  \ocite{BP57,P60,BS80} who constructed a
recursive subtraction scheme --- the $R$-operation --- to
remove  ultraviolet (UV) divergences
from a given Feynman integral in
a way compatible with adding local counterterms to the
Lagrangian.  Unfortunately their proof of the main theorem of
the renormalization theory, ---  the fact that the
$R$-operation does subtract {\it {all}} infinities --- had
included an intermediate statement which in fact was not
true.  This has been corrected by Hepp  \ocite{Hepp} and
hence the theorem is known as BPH (a simpler version of the
proof have been presented by Anikin, Zavialov and
Polivanov \ocite{AZP73}).

A major step in elaborating the discussed approach to
renormalization has been made by Zimmermann \ocite{Z69,Z70,Z73a,Z73b,Z76}.
In particular, he has introduced the concept of
oversubtractions and developed a graph-theoretical
forest technique to
disentangle the complicated recurrence structure of the
$R$-operation with oversubtractions. (Note that the forest
formula for the $R$-operation without oversubtractions was
first derived by Zavialov and Stepanov in a somewhat
disguised form \ocite{ZS65}).  The forest technique and
oversubtractions
form the basis for the normal product method which has been
of great use in treating theories with massless
particles \ocite{Lowen75a,Lowen75b,Lowen75c,Lowen76c},
 in
deriving rigorously  Wilson expansion  \ocite{Z70,Z73b,Clark74},
and in proving  various general relations between
renormalized Green functions meet (Zimmermann
 identities \ocite{Z70,Z73a,Lowen72,Lowen73},
 renormalization group
equations \ocite{Lowen71a}, the Quantum Action
Principle \ocite{Lowen71a,Lam7273,Lowen76b}, etc.).
Further  important development of the method has
been made by Anikin, Polivanov  and Zavialov 
 \ocite{AZ76,AZP77,An82,Za89}.

A notorious problem of the renormalization theory is the
choice of the ultraviolet cut-off and of the renormalization
scheme.  Though in perturbation theory one can manage to
avoid any cut-off (by using the
so-called regulator-free formalism which
proves to especially useful in rigorous study of
supersymmetric theories \ocite{Klaus86}) in general   there
is   no a preferred  choice of the renormalization
scheme.  On the other hand, a suitable choice of
regularization and of the renormalization prescription can
facilitate doing field theory considerably.

In practice, the dimensional
regularization \ocite{HV72,Bollini,BM77a,BM77b,BM77c}
(DR) has become a very useful tool in perturbative treatment of
various field theories, including the non-abelian gauge ones. In
particular, the DR is used  to perform virtually all
complicated perturbative calculations. There are at least two
good reasons for doing so, in addition to its explicit gauge
invariance. The first one is that
within the DR the divergent FI's
can be treated, in many aspects, as if
they were convergent. In other words,  such
operations as cancellation of identical factors
in the numerator and denominator of the  integrand,
(formal) integration by parts, and  replacement of the
integral of a sum by the sum of
the corresponding integrals are well-defined
(a more complete list is given in \ocite{me83c}).
Another very useful property  of the DR is its ability to
regularize simultaneously both the UV and the infrared (IR)
divergences by transforming these into poles in
$\ep=(D_0-D)/2$, where $D_0>0$ is the integer dimension of
space-time while $D$ is the running dimension.

As for the choice of the renormalization prescription, it is the
minimal subtraction (MS) scheme \ocite{Hooft73}
along with its straight\-forward
modifications such as the $\overline{MS}$-scheme \ocite{MSbar}
 and the $G$-scheme
 \ocite{me80}
that prove to be very convenient both  for  calculation and
for phenomenological applications.

A remarkable feature of the MS-scheme is the fact that in its
framework all UV counterterms (or, equivalently, UV
renormalization constants) are polynomial  {\it both} in
momenta (which must be the case for every meaningful
renormalization prescription) {\it and} in  masses \ocite{JCC75}.
It is this
property, along with the useful features  listed above,
that gives the DR its calculational power in such problems
as evaluation of various renormalization group
functions.  Indeed, the overall divergence of a log-divergent
Feynman integral (i.e. the divergence remaining after
minimally subtraction of all its subdivergences) must be a
polynomial in $\ep^{-1}$, with purely numerical  coefficients
{\it without} any dependence on dimensional
parameters. Thus, when computing the UV renormalization constants
one is free to perform arbitrary rearrangements of masses and
external momenta, e.g to nullify some of them {\it provided}
this does not lead to IR divergences.
This observation , first made
in ref. \ocite{Vla80a}  (see also \ocite{me80,CK82}) has since
then been used repeatedly in a
number of important calculations.
Among the latter are the analytical evaluation of the
$\alpha_s^2$-correction to the total cross-section of the
$e^+\,e^-$ process \ocite{me79} and the calculation of  the
$\beta$-functions for QCD  and N=4 supersymmetric Yang-Mills
theory \ocite{Vla80b,Vla80c} both made at the three-loop level.

However, the condition of that the IR divergences do not appear
restricts
considerably our calculational abilities, since for complicated
FI's this requirement prevents one from reducing a given FI to a
simpler one (see e.g. ref. \ocite{me82a} and Sect.4).

The $R^*$-operation --- a generalization of the
$R$-operation for
subtracting both UV and IR divergences --- was invented \ocite{me82a}
just to
overcome
this difficulty and solved the problem by allowing for arbitrary
rearrangements of masses and external momenta \ocite{me82a,me84}.
This, in turn,
has greatly extended the class of problems amenable to
analytical solution. It is sufficient to say
that it was the use of the
$R^*$-operation that has enabled one to   evaluate
analytically  the $\beta$-function at the {\it five} loop level for
two theories: the $\phi^4$ model \ocite{me83a,Kaz83} and the
supersymmetric two-dimensional sigma-model \ocite{Kaz87}.

Another field of interest where the $R^*$-operation
has been working successfully is the investigation of
various short-distance expansions and  heavy mass decoupling
 \ocite{me83b,T83a,me87a,GorLar87,me88a,me88b,Smi88,Smi89,Smi91}.

The so-called $R^{-1}$-operation --- the inverted
usual $R$-operation ---  was first introduced in \ocite{T86,T88}
in the study of asymptotic expansions
of minimally subtracted FI's.
An explicit resolution of the
involved recursion structure of the $\Rm$-operation  was
found in \ocite{me87b,me89b} and then applied to construct
a rigorous  algebraic derivation    of the
renormalization group equations in the $MS$-scheme \ocite{me89b}.

The present work is aimed at making a regular and uniform
treatment of the combinatorics
of $R$-, $\Rm$-,   and $R^*$- operations and of
the generic large-mass and/or
momentum asymptotic expansions of euclidean FI's.

The outline of this  work is as follows.
In the next section we introduce definitions and notations for
Feynman graphs, integrals and the generic subtraction operation.
Section 3 is devoted mainly to the derivation of a
forest representation
of the $\Rm$-operation and to studying its implications
for the BPHZ renormalization
formalism.  We  show that the $\Rm$-operation reveals
the underlying simple structure of the BPHZ renormalization
and is a natural tool for constructing general conversion formulas
connecting two different renormalization versions of a given FI.
As an example we give a simple derivation of  the Zimmermann
identities and the conversion formulas from the MS-scheme to
the  momentum subtraction and vice versa.

In Sect.4 we develop the theory of the $R^*$-operation
in a   general situation where the Feynman integrals
to be renormalized may (i) have
their external momenta put under an arbitrary number of linear
constraints  and (ii) may be formally expanded in
some external momenta and/or masses\foot{
That is one first  expands  the corresponding {\it integrands}
and then dimensionally regularizes  \mbox{resulting FI's.}}.
It is shown that
in the $MS$-scheme  the $R^*$-operation may be
naturally expressed
through the  usual $R$-operation and its inverse ---  the
$\Rm$-operation.
Then, as a result of this  representation we
get a simple and regular
algorithm for evaluating the UV counterterms of
arbitrary dimensionally
regularized FI's, including the ones appearing
in the two-dimensional field models.

Section 5 is devoted to combinatorics of the
large momentum and/or heavy mass expansions of euclidean,
minimally renormalized FI's in a very general setting.
The integrals under consideration may be of arbitrary form,
including the non-scalar and (completely or partially)
massless ones, and with their external momenta may be
subjected to any number of linear constraints.
This also includes the case
where these integrals are considered as being
formally expanded
in some of
their external momenta and/or masses.

It should be stressed that the above integrals form a
natural class for studying large momentum and/or
heavy mass behavior. This is due to the fact that the
resulting asymptotic series can be expressed exclusively in terms of
the FI's from the same class.

Our starting point is  Theorem  8  of subsect. 5.1 that
describes the  complete asymptotic expansion of FI's
of the form specified above at $\rho\to0$,  in which some of
their external momenta and/or masses are scaled by
$\rho^{-1}$.  The expansion is presented in a concise form
convenient for practical calculations: in terms of  the
(again dimensionally regularized!) FI's for some subgraphs
and reduced graphs.  Once the  expansion of
regularized FI's at $\rho\to 0$  is at hand it becomes a purely
combinatorial problem to derive  the corresponding expansion
for minimally renormalized FI's or Green functions. The
complete solution of the problem  given in subsect. 5.2
naturally leads to the appearance of the $R^*$-operation in
the resulting, explicitly  finite expansion.  The section
ends with a simple combinatorial proof of the main
theorem of the $R^*$-operation  theory  --- of the fact
that this does subtract all divergences from the
generic dimensionally regularized FI.

Section 6 contains a comparative discussion of the relevant
results of previous  studies. Finally, in  Sect.7
we present our main conclusions together with a brief
outline of some  of the problems awaiting their solution in
the developed framework.

\chap{STRUCTURE OF FEYNMAN GRAPHS AND INTEGRALS}
In this section we briefly recall basic graph-theoretical
notations relevant to (Feynman) graphs and integrals.
The material is partially taken from
refs. \ocite{Sp76,CK82,me89a}.
We also discuss the definition and main properties
of the $c$-operation --- a generalization of the standard
ultraviolet counterterm operation $\D_U$ ---
 which will be in constant use in the following.

\sect{Basic graph-theoretical notations and definitions }
A graph $\G$ is a set of lines and vertices which can be
associated
with a Feynman integral (FI) --- a term in the perturbation
expansion.
The collections of internal lines, vertices and external lines
will be
denoted as $\LG$, $\VG$ and $\EG$, while
$L_\G=\vert\LG{\vert}$, $V_\G=\vert\VG {\vert}$ and
$E_\G=\vert{\EG} {\vert}$ will stand for the numbers
of elements in
each set. Every internal line
$l\in\LG$ has two
(possibly coinciding) vertices incident to,
it, viz., the initial
vertex $\pi_-(l)\in\VG$ and the final vertex
$\pi_+(l)\in\VG$. Every external line
$l\in\EG$ has with one vertex,
$\pi(l)\in\VG$, incident to it, and a sign factor
$f_{\G}(l)=\pm1$.
If
$f_{\G}(l)=+1(-1)$, the corresponding external momentum
is  supposed to flow in(out) the vertex $v=\pi(l)$.

The empty graph
$\G_\emptyset$ is the unique graph with
${\cal  L}_\emptyset =\emptyset$,
${\cal  V}_\emptyset =\emptyset$ and
${\cal  E}_\emptyset =\emptyset$.

Let $l\in\LG$. A subgraph $\G^\prime=\G \backslash l$ of
$\G$ is obtained via deleting $l$ from $\LG$ and adding two
new external lines $l_-$ and $l_+$ produced by "cutting" the
line $l$ and incident to the vertices $\pi_-(l)$ and
$\pi_+(l)$, respectively.  The mappings  $\pi'_{\pm}$ and
$\pi'(l)$ and the function $f_{\G'}$ are defined  in the
natural way, \foot{ That is as restrictions of $\pi_{\pm}$,
$\pi$, and  $f_{\G}$ onto ${\cal  L}'={\cal  L}\b l$, and ${\cal  V}'={\cal  V}$
respectively.} with $\pi'(l_{\pm})=\pi_{\pm}(l)$, and
$f_{\G'}(l_\pm)=\pm1$.  A subgraph $\G\backslash{\cal  L}^\prime$
is defined for any set ${\cal  L}^\prime\subset{\cal  L}$  by repeated
applications of this prescription. Every subgraph $\g$ of
$\G$ is (unambiguously) determined by a pair of subsets
$\Lg\subset\LG$ and $\Vg\subset\VG$ ($\Vg$ is understood to
comprise all the vertices incident to the lines from $\Lg$
).  It is obtained from the subgraph $\G^\prime
=\G\backslash (\LG\backslash\Lg)$ by throwing away all the
isolated vertices of $\G^\prime$ not belonging to $\Vg$.
Thus, the external lines incident to a vertex $v$ of $\g$
are the original external lines together with some
"fragments" of internal lines from $\LG$.  Sometimes we
shall write ${\cal  L}$, ${\cal  E}$, ${\cal  V}$ and ${\cal  L}(\g)$, ${\cal  E}(\g)$,
${\cal  V}(\g)$ instead of $\LG$, $\VG$,  $\EG$ and ${\cal  L}_\g$,
${\cal  E}_\g$, ${\cal  V}_\g$ respectively.  
Every subgraph $\g$ of $\G$, except for the $\G$ itself
is said   to be a (proper) subgraph of $\G$.
 If $\g$ is a (proper)
subgraph of $\G$, we shall write
$\g\subseteq\G(\g\subset\G)$. A graph $\g$ is trivial if
$\Lg=\emptyset$. If $v\in{\cal  V}$ then $\dot v$ will stand for
the (unique) trivial subgraph of $\G$ with ${\cal  V}_ {\dot v} = v
$.

The number of {\it c-components} (that is  of the maximal
connected subgraphs of $\G$)  and
independent loops (circles)
of $\G$ will be denoted as $c(\G)$ or, equivalently,
$c_{{}_\G}$ and
${\cal  N}(\G)=L_\G - V_\G + c_{{}_\G}$   respectively.

If $\g$ and $h$ are two subgraphs of $\G$, their {\it union}
$\delta=\g\cup h$
is the subgraph with ${\cal  L}_{\delta}=\Lg\cup{\cal  L}_{a}$ and
${\cal  V}_{\delta}=\Vg\cup {\cal V}_h$;
their {\it intersection} $\g\cap h$
is defined in a similar way. $\g$ and $h$ are called {\it disjoint}
if $\g\cap h=\G_\emptyset$.

We skip over the well-known definitions of a {\it connected} graph
and of a {\it one-particle-irreducible} (1PI) graph. Given a connected
graph $\g\subset\G$, the {\it reduced} graph $\G/\g$ is obtained
by reducing $\g$ to a single vertex, $v_\g$. If $\g$ is a
disconnected subgraph than $\G/\g$ is produced by reducing
each of its  c-components $\g_i$ into a single vertex, $v_{\g_i}$.

A (proper) {\it spinney} $S$ of a graph $\G$ is a pairwise
disjoint family of non-empty
1PI (proper) subgraphs
of $\G$. A (proper) {\it wood} $W\{\G\}$\ ($\bar W\{\G\}$)
is the collection of all (proper) spinneys of $G$. A (proper)
{\it forest} $F$\,($\bar F$) is a set of non-empty
1PI subgraphs of $\G$
such that if $\g$,$\g^\prime\in F$  then  either
$\g\cap\g^\prime=\G_\emptyset$, or  $\g\subset\g^\prime$
or $\g^\prime\subset\g$. The collection of all (proper)
forests of $\G$ will be denoted as $F\{\G\}\ ($($\bar F\{\G\}$).
Given a forest $F$, we shall denote by $\vert F\vert$ and
$(F)_{max}$,  respectively,  the number of elements in $F$ and
the (unique)
maximal spinney $S$ such that $S\subseteq F$.
The empty spinney $S_\emptyset$, i.e. the unique spinney without
elements, belongs  to both  $W\{\G\}$ and $F\{\G\}$.

In what follows
we shall repeatedly deal with various relations between forests.
Being a particular example of a forest,
a spinney (or even a single graph)
may also take part in
these relations. Let $F$ and $F^\prime$  be two forests.
We shall define the relations between these forests in terms of
their members $\g\in F$ and $\g^\prime\in F^\prime$.
$F\sim F^\prime$ if for all $\g$ and $\g^\prime$
\  $\g\cap\g^\prime=\G_\emptyset$; $F\geq (>) F^\prime$
if every $\g^\prime$ is a (proper) subgraph of an element
from $F$; $F\gi F^\prime$ if for every $\g^\prime$
either $F>\g^\prime$ or $F\sim \g^\prime$; $F\succ F^\prime$
if $F\gi \g^\prime$ and there is no such element
 $\g\in F$ that $\g\sim F^\prime$. For every nonempty forest
$F$ one has  $S_\emptyset\sim F$, $F>S_\emptyset$,\  $F\gi S_\emptyset$,
\ $S_\emptyset \gi F$ and $S_\emptyset\succ F$.

If $S$ is a spinney of $\G$, then $\G/S$ will stand
for the graph obtained by reducing every non-trivial
element of $S$.
By $S\!\downarrow\!\g$ we shall mean a spinney formed
by all the elements of $S$ which are subgraphs of $\g$.
If $S_1$ and $S_2$ are two spinneys of $\G$ such that
$S_1\gi S_2$ or $S_1\ge S_2$,
than the family $S_1/S_2$ is defined according to
\beq S_1/S_2=\{\g/(S_2\!\downarrow\!\g)\vert\g\in S_1\}\nnb \eeq
and is to be, naturally, interpreted as a spinney from
$W\{\G/S_2\}$.

Finally, we shall denote the collection of graphs
$\{\g\vert\g=\delta/S,\ \delta\subseteq\G,\ S\in W\{\delta\}\}$
as ${\cal  P}\{\G\}$.

\sect{Dimensionally regularized Feynman integrals}
Having recalled the graph-theoretical background,
we turn to the second part of the notion of
the Feynman diagram --- the Feynman integral.
Consider an unrenormalized Feynman amplitude
$\bar F_\G({\bf  q} ,{\bf  m})$
corresponding to a connected
graph $\G$ with ${\bf  q}=\{q_l\vert l\in{\cal  E}\}$ and
${\bf  m}=\{m_l\vert l\in{\cal  L}\}$  standing for its
external momenta and masses, respectively. Within the dimensional
regularization it can be formally represented as\foot{We use this formal presentation
only as a convenient substitution for  rigorous
definitions of refs. \ocite{BM77a,me83c,CM83}.
The main role of the former is to
help one to formulate concise definitions of
various operations with dimensionally regularized FI's.}
\beq
{{\bar F_\G({\bf  q},{\bf  m},\mu,\ep)=\delta_\mu (\dsp \sum_
{l\in{\cal  E}}f_{{}_\G}(l) q_l)<\G>({\bf  q},{\bf  m},\mu,\ep),}
\atop\displaystyle
 <\G>=\int
 Y_\G({\bf
   q},{\bf
   k
 })\,d_\mu{\bf
   k
 },}
\EQN{221}
\eeq
where
\beq 
Y_\G=\prod_{v\in{\cal  V}}{{P}}_v
({\bf  q}^{\dot v})
\prod_{l\in{\cal  L}}
{{D}}_{l}(p_l),\
{ d_\mu}{\bf  k}=\prod_
 {i=1}^{{\cal N}(\G)}
d_\mu k_i,\
d_\mu k_i=
 {{\mu^{2 {\ep}}}\over{(2 \pi)^D}}dk_i,
\nnb
\eeq
and  $D_l(p_l)=P_l(p_l)/(m_l^2 + p_l^2)$, with
$P_l$ being a  polynomial of degree $a_l$.
The momentum $p_l$ is a linear combination of the
loop momenta ${\bf  k}=\{k_1,\dots k_{{\cal N}_\G}\}$
and of the external
momenta flowing through the internal line $l$.
$P_{v}
({\bf  q}^{\dot v})$ is a  polynomial of degree
$a_v$ in momenta
${\bf  q}^{\dot v}=\{q_l^{\dot v}\vert l\in{\cal  E}_{\dot v}\}$ ---
 the set of external momenta of the (trivial) subgraph
$\dot v\subset\G$. By definition, $q_l^{\dot v}=
q_l$ if $l\in{\cal  E}_\G\cap{\cal  E}_{\dot v}$, and
$q_l^{\dot v}=p_{l^\prime}$ if $l\in{\cal  E}_{\dot v}$,
$l=l^\prime_{\pm}$ .
$D=D_0-2\ep$ is the running space-time dimension,
with the  (positive) integer $D_0$ being the physical
one.

We
have introduced the mass $\mu$ to preserve
the correct dimension of $<\G>$; after minimally subtracting
the UV poles in $\ep$, $\mu$ will serve as the MS-scheme
renormalization parameter. Finally, every polynomial
$P_l$ (and $P_v$) should be considered as a member of
the formal algebra of $D$-dimensional covariants
(which includes the $\g$-matrices and
the metric tensor in addition
to the $D$-dimensional vectors in the case of theories
with fermion fields). We shall assume that every polynomial
that may serve as a particular example of $P_l$ (or
$P_v$) is written in the normal form \ocite{BM77a} and,
thus, does not contain any explicit dependence on $\ep$.
The index of UV divergence of the  FI $<\G>$ reads
\beq \omega(\G)= D_{0}{\cal  N}_{\G} -2L_{\G}+\sum_{l\in{\cal  L}_{\G}}a_{l}
+\sum_{v\in{\cal  V}_{\G}}a_{v}.\nnb \eeq

Let us denote a formal algebra
formed by polynomials  in $q_l \in{\bf  q}$
as
$A({\bf  q} ,\ep)$; 
the coefficients of
these polynomials may depend meromorphically on $\ep$. By
$\{{\bf{e}}_i({\bf  q} )\}$ we shall mean
a monomial basis in the
algebra, so that every element $e\in
A({\bf  q} ,\ep)$
can be uniquely presented in the form
\beq e=\sum_i\xi_i(\ep){ e}_i({\bf  q}^\g),
\nnb \eeq
with all its dependence on $\ep$ concentrated in the coefficients
$\{\xi_i\}$.

For a given  a connected subgraph
$\g\subseteq\G$, the corresponding Feynman subintegral
$<\g>$ is the FI defined by \re{221}, with the polynomials
$P _v^\g$
and
$P_l^\g$
being equal to
$P_v $
and
$P_l $,
respectively. If S is a spinney, then we shall denote the FI
corresponding to the (disconnected) graph
${\bigcup_{\g\in S}}\g\ $ as $\dsp<S>=\prod_{\g\in S}<\g>.$

Now let $S\in W\{\G\}$ and let
$\{{ e}_{i_\g}({\bf  q}^\g)\vert\g\in S\}$
be a collection of basic monomials. Proceeding in the same
manner we construct the FI $<\G/S>^{\bf i}$
with ${\bf i}=\{i_\g\vert\g\in S\}$ from propagators and
vertex factors of $<\G>$ ,with the understanding  that
every element ${e}_{i_\g}$ serves as
the  vertex polynomial
$P_{v_\g}^{\G/S}({\bf  q}^{\dot v_\g})$
(missing in the initial FI $<F>$ ).
Finally, if
$\ \{P_\g\vert\g\in S\}\ $
is a collection of polynomials such that
\beq P_\g\in A({\bf  q}^\g,\ep),\ \ %
 P_\g=\sum_{i_\g}{\xi}_{i_\g}(\ep){ e}_{i_\g}({\bf  q}^\g),
\nnb \eeq
then we define the FI
$(\prod_\g P_\g)\ast<\G/S>$ as the following
multifold series of FI's
\beq \sum_{\bf i}(\prod_{\g\in S}{\xi}_{i_\g})<\G/S>^{\bf i}.
\nnb \eeq

\sect{$c$-operation and its properties}
Let us suppose that we are provided with a rule, $\Delta$,
which associates a polynomial from
$A({\bf  q}^\g,\ep)$ to every FI $<\g>$ with $\g$ being
a 1PI graph:
\beq \D<\g>=\sum_i\xi_i(<\g>,\ep)\ {\bf  e}_i({\bf  q}^\g),\EQN{231}\eeq
where $\xi_i$\, is a meromorhic function of $\ep$.
Unless otherwise is explicitly stated
we shall also demand  that $\D<\g>=0$ if $\g$ is
a trivial graph.

If $\g\subseteq\G$ then the action of the
$c$(counterterm)-operation $\D(\g)$
corresponding to the rule $\D$
on a FI $<\G>$ is defined as follows
\beq \D(\g)<\G>=\sum_i\xi_i(<\g>,\ep)({\bf e}_i\ast<\G/\g>)\nnb \eeq
  
or, equivalently,
\beq \D(\g)<\G>=\D<\g>\ast<\G/\g>.\nnb \eeq
  
Next, let $S$ be a spinney of $\G$ and $<\G/S>$ be an arbitrary
FI corresponding to the graph $<\G/S>$. We define
\beq
\D(\g)<\G/S>=\left\{\begin{array}{l}
<\g/(S\!\!\downarrow\!\g)>  \ast<\G/(S\cup\g)_%
{max}>, \ \   \mbox{if} \ \mbox{$\G\supseteq\g\gi S$;} \\
0,\ \ \ \ \  \mbox{otherwise.}
\end{array}
\right.
\nnb \eeq
\ice{
\beq \D(\g)<\G/S>=\cases {\D<\g/S\!\do\!\g>\ast<\G/(S\cup\g)_
{max}>,&if $\G\supseteq\g\gi S$;\cr
0,&otherwise.\cr}\nnb \eeq
}
Thus, if $h\subset\g\subseteq\G$
and $\D$ and $\D'$ are a pair of (possibly identical)
$c$-operations then
\beq \D(h)\D'(\g)=
\D(\g)\D(\g)=
\D'(h)\D'(h)=0,\nnb \eeq
and
\beq \D'(\g)\D(h)<\G>=\sum_i\xi_i(<h>,\ep)\D'(\g)<\G/h>^i.\nnb \eeq
Moreover, if $\g,h\subseteq\G$ and $h\cap\g=\G_\emptyset$
then
\beq \D'(h)\D(\g)=
\D(\g)\D'(h)\nnb \eeq
and
\beq \D(\g)\D'(h)=\sum_{i,i^\prime}\xi_i'(<\g>,\ep)\xi_{i^\prime}'
(<h>,\ep)<\G/(\g\cup h)>^{i,i^\prime}.\nnb \eeq

If $S\in W\{\G\}$, then  we put
\beq \D(S)=\prod_{\g\in S}\D(\g),\EQN{232}\eeq
with
\beq \D(S)<\G>=\D<S>\ast<\G/S>\ \  \
\hbox{\rm and}\ \ \ \D<S>=\prod_{\g\in S}
(\D<\g>).\nnb \eeq

Similarly, if  $F$ is a forest, then
\beq \D(F)=\prod_{\g\in F}\D(\g),\EQN{233}\eeq
where it is understood that in the product in the r.h.s. of
\re{233} the order is determined by
the $c$-operation $\D(\g)$ for smaller
graphs acting first (on the right).
As a consequence of \re{232}  we get
that for the empty spinney $S_\emptyset$
\beq \D(S_\emptyset)=1\nnb \eeq

It is worth  noting that as for their algebraic properties
$c$-operations are, in fact, identical to the subtraction
operators widely used in the renormalization theory (see,
e.g. \ocite{Za89}).

\chap{$R$- AND $R^{-1}$- OPERATIONS}
In this section we first give a precise definition of
the  $R$- operation corresponding
to a given $c$-operation and  elaborate on its inverse --- the
$R^{-1}$- operation ---, which will prove to be a very
convenient tool in treating both the $R^*$-operation
and the asymptotic expansions of Feynman integrals.
Then
we consider the particular versions of these operations
adopted to the MS-scheme and reveal the deep connection between
the $R^{-1}$-operation  and the BPHZ renormalization and
finally derive
 general conversion  formulas
connecting the FI's renormalized according to different
renormalization
prescriptions, viz. by the minimal subtraction and
by the BPHZ momentum  subtraction.

\sect{$R$-operation}

Let $\D$ be an arbitrary $c$-operation  and $\G$ be a
(not necessarily
connected) graph.
According to the  notation
 developed above
the  corresponding $R$-operation reads as
\beq R(\G)=\sum_{S\in \Wu\{\G\}}\D(S),\EQN{r1}\eeq
where the wood\foot{The subscript $U$ is to recall that
the genuine $U$ltraviolet subdivergences are connected
with 1PI, non-trivial subgraphs.} $\Wu\{\G\}$ comprises only
such spinneys from $W\{\G\}$ that do not include trivial
subgraphs of $\G$ as their members.
In addition, it is also convenient to define
\beq {}'\!R(\G)=\sum_{S\in \Wu\{\G\}}^{S\not=\G}\D(S)\equiv
R(\G)- \Delta(\G),
\nnb 
\eeq
and write $R\!<\G>$ and ${}'\!R\!<\G>$ instead of
$R(\G)\!<\G>$ and ${}'\!R(\G)\!<\G>$, respectively.
\eol


 {\bf
\noindent 
  Theorem
  1}
{\em
\begin{description}

\item{(a)} If $\G$ is a trivial graph, then
\beq R(\G)=\D(S_\emptyset)=1.\EQN{r2}\eeq

\item{(b)} If $\D_1$ and $\D_2$ are two (not necessarily different)
$c$-operations, then
\beq R_1(\G)R_2(\G)=\sum_{S\in \Wu\{\G\}}
\prod_{\g\in S }
(\D_1(\g){}'\!R_2(\g)+\D_2(\g)).\EQN{r3}\eeq

\item{(c)} Let $\g$ be a graph of the form $\g=\delta/ \Phi$
with $\delta\subset\G$ and $\Phi$ being a spinney of $\G$
such that $\delta>\Phi$. Then there holds the following
chain of equations:
\beq R(\G)\!<\delta/\Phi>=\sum_{S \in \Wu\{\g\}}\D(S)
\!<\delta/\Phi>= R\!<\g>.\EQN{r4}\eeq

\end{description}
\label{Rproperties}


\noindent
{\bf Proof. }}

\begin{description}

\item{(a)} Follows directly from the definition of the  $c$-operation.

\item{(b)} Indeed, due to the a properties of
the product of $c$-operations
 discussed above
\beq R_1(\G)R_2(\G)=\sum_{S'\in \Wu\{\G\}}
\sum_{S\in \Wu\{\G\}}^{S'\gi S}\D_1(S')\D_2(S).\EQN{r5}\eeq
  
Further, if a spinney $S$ obeys the relation $S'\gi S$
then it can be unambiguously decomposed into two "subspinneys"
$S_a$ and $S_b$ such that $S=S_a\cup S_b$,\
$S_a<S'$ and
$S_b\sim S'$. This allows one to transform
the second sum over $S$ in the r.h.s. of
\re{r5} as follows:
\beq \sum_{S_a<S^\prime}
\sum_{S_b\sim S^ \prime }\D_1(S')\D_2(S_a)\D_2(S_b)=
\Bigl(\prod_{\g^\prime\in S^\prime}
\D_1(\g^\prime){}'\!R_2(\g^\prime)\Bigr)
\Bigl(\sum_{S_b\sim S'}
\D_2(S_b)\Bigr).\nnb \eeq
  
Using this relation and going to the summation
over $S=S'\cup S_b$ and $S_b$,
one readily obtains \re{r3}.

\item{(c)} As a direct consequence of the
properties of $c$-operations,
one gets
\beq R(\G)\!<\g>=\sum_{\Phi\li S\in \Wu\{\delta\}}
\D(S/\Phi)\!<\delta/(\Phi\cup S)_{max}>.\nnb \eeq
  
On the other hand, the mapping
\beq \{S\vert\Phi\li S\in \Wu\{\delta\}\}
{\ {}^{\phantom{aaaa}r}{\!\!\!\!\!\!\!\longrightarrow\ }}
\Wu\{\delta/\Phi\} \nnb \eeq
  
with $r(S)=S/\Phi$ provides a one-to-one correspondence
between these two sets, whence \re{r4} follows.
\end{description}

\sect{$R^{-1}$-operation}
Let us define
\beq R^{-1}(\G)=\sum_{S\in \Wu\{\G\}}\D^{-1}(S),\EQN{rm1}\eeq
where $\D^{-1}$ is a $c$-operation\foot{$\Dm$ is {\it not}
an inverted $\D$!}.
It was shown in refs. \ocite{T86,T88}
that the $c$-operation $\ \Dm$
is unambiguously determined by demanding that
\beq R(\G)\Rm(\G)=1\EQN{rm2} \eeq
and
\beq 
\Rm(\G)R(\G)=1\EQN{rm3}.
\eeq
We are going to find an explicit
representation for the $\Rm$-operation
in terms of the $\D$-operation. We begin
by proving that \re{rm2}
 is equivalent
to the following equation
\beq \D(\g)\Rm(\g)=-\Dm(\g)\EQN{rm4}\eeq
provided the latter holds for any 1PI graph $\g$.
Indeed, given \re{rm4}, \re{rm2}
is derived by a straightforward application of \re{r3}
\beq R(\G)\Rm(\G)=1+\sum_{
\emptyset\not=S\in \Wu\{\G\}}\prod_
{\g\in S}(\D(\g)\Rm(\g)+\Dm(\g))\EQN{rm5}\eeq
  
On the other hand, if \re{rm2} holds then \re{rm4}
 can be  proved  trivially
by   considering \re{rm5} with $\G=\g$ and using induction in the loop
number
${\cal N}_\g$.

The identity \re{rm4} is a convenient starting
point for constructing $\Dm$.
Let us rewrite it as follows
\beq \Dm(\G)=-\D(\G)-\sum_{\emptyset\not=S\in \Wu\{\G\}}
\D(\G)\Dm(S).\EQN{rm6}\eeq
Due to the identity $\D(\G)\Dm(\G)=0$,$\ $ \re{rm6}
expresses the $\Dm$-operation in terms of the $\D$ and
$\Dm$-operations, the latter appearing only for graphs with
their loop number less than ${\cal N}_\G$.  The next step is
to use the same equation (with $\G$ substituted by $\g$ with
$\g\in S$) for every factor $\Dm(\g)$ in the r.h.s of
\re{rm6}.  This recursion process will stop when there
remain only one-loop diagrams as arguments of
the $\Dm$-operations. The resulting sum may be further
simplified by collecting similar terms and the final result
reads
\beq \Dm(\G)=-\D(\G)\sum_{F\in \bar \Fu\{\G\}}(-)^{ \vert
F\vert}\D(F)\EQN{rm7}
{},
\eeq
where the the collections of forests $\Fu\{\G\}$ comprises
only such forests from $F\{\G\}$ that do not include
trivial graphs as their elements.

{\bf  Proof.} By induction in the loop number ${\cal N}$.
The statement is obviously true at the one-loop level.
Assuming its validity for ${\cal N}(\G)\leq n$, we transform
the relation \re{rm6} with ${\cal N}(\G)=n+1$
as follows:
\beq \Dm(\G)=-\D(\G)\sum_{S\in \Wu\{\G\}}\prod_{\g\in S}\left
\{-\D(\g)\sum_{ F\in{\bar{F}}\{\g\}}(-)^{\vert F\vert}
\D(F)\right\},\nnb \eeq
  
whence \re{rm7} follows immediately.  Note that the representation
\re{rm7} leads directly to {\it the forest formula }
for the $\Rm$-operation, viz.
\beq R^{-1}(\G)=\sum_{F\in \Fu\{\G\}}
(-1)^{\vert F\vert}
\D(F).\EQN{rm8}\eeq
  
As we shall see later on, the  remarkable similarity
between \re{rm8} and the forest formula for the BP
$R$-operation (see  subsection 3.4) is by no means
an accidental coincidence but, rather, reflects the deep
relation of the $\Rm$-operation to the BPHZ formalism.

\sect{$R$-  and $\Rm$-operations within the MS-scheme}

The main application of the $R$-operation is to make a given
FI finite at  $\ep\rightarrow 0$ via subtracting from it the
UV divergences in a way compatible with the general
principles of quantum field theory (see, e.g. refs. \ocite{BS80,Za89}).  
Any particular version of renormalization
scheme can be obtained by making a proper choice of the
corresponding $c$-operation. A very convenient choice,
inherently connected with dimensional regularization, is the
minimal subtraction scheme, with its  main virtue of
respecting formal symmetries\foot{With notorious exceptions
 being symmetries evoking chiral transformations.} of
 dimensionally regularized FI's and, hence, dimensionally
regularized Green functions.

Let $A({\bf  q}^\g,{\bf  m}^\g)$ denote the formal algebra
$A({\bf  q}^\g)$  extended by allowing for every mass from
${\bf  m}^\g$ to act as an extra generator. In other
words, every ${\bf  e}_i\in A({\bf  q}^\g,{\bf  m}^\g)$
is, in fact, a basis vector from
$A({\bf  q}^\g)$ multiplied by a monomial in masses from ${\bf  m}^\g$.
The $R$-operation in the MS-scheme, $\Ru$,
which,  when applied to a FI $\!<\G>$, converts the latter
into the corresponding minimally subtracted FI, reads%
 \ocite{JCC75,Vla80a,me80}
\beq \Ru(\G)=\sum_{S\in \Wu\{\G\}}\Du(S).\EQN{ms1}\eeq
Here the $c$-operation $\Du(\g)$ is supposed to evaluate
the ultraviolet MS-counterterm corresponding the
overall UV divergence of the FI $<\g>$.
Considering the linearity of the
 $\Ru$-operation \ocite{JCC75}, we may assume, without
essential loss of generality, that polynomials $P_v$ and
$P_l$ are homogeneous. This allows us to present
the action of the c-operation $\Du$ on a FI $<\G>$ in
the following form
\beq \Du\!<\g>=\sum_{i}Z_{i}(\!<\g>,\ep)
{\bf  e}_{i}({\bf  q}^{\g},{\bf  m}^{\g}),
\EQN{ms2}\eeq
where $\{{\bf  e}_{i}({\bf  q}^{\g},{\bf  m}^{\g})\}$ are the basis vectors
of mass dimension  $\omega(\g)$ from  $A({\bf  q}^\g,{\bf  m}^\g)$.
The dimensionless renormalization constants $\{Z_i\}$
are polynomials in $1/\ep$ such that
\beq K\,Z_{i}=Z_i\EQN{ms2a}
{},
\eeq
where $K\,f(\ep)$ stands for the singular part of
the Laurent expansion of $f(\ep)$
in $\ep$ near $\ep=0$.

The main theorem of the renormalization
theory in the case of
the MS-scheme can be formulated as follows \ocite{BM77a}:

{\bf Theorem 2} {\it There exists a unique
choice of the
renormalization constants $\{Z_i\}$
fulfilling the minimality
restriction \re{ms2a} and such that  the $\Ru$-operation
makes  arbitrary
{\rm (}infrared convergent\foot{{\rm Any euclidean FI
without massless lines is
infrared convergent; the FI's that
include massless lines
are discussed in the next section.}}{\rm)}
FI finite  in the limit $\ep\rightarrow 0$.}

The finiteness of $\Ru\!<\G>$ at $\ep\to0$ means that
\beq \Du\!<\g>=-K{}'\!\Ru\!<\g>.\nnb \eeq
  
This  equation may be used to define
the UV counterterms and, thereby, the $\Ru$-operation
for any FI's including the IR divergent ones.
This is achieved by (i) introducing
an auxiliary non-zero mass $\mu_0$ into every
massless propagator to ensure suppression of all
the IR divergences; (ii) computing $\Du\!<\g>
({\bf  q}^\g,{\bf  m}^\g,\mu_0,\mu,\ep)$ and (iii)
setting $\mu_0=0$ in the final result. Being quite
correct this {\it a brute force  }
prescription is surely not to be considered
as a practical one; there exist much more
subtle tricks that are more convenient for
practical calculations (see refs.
\ocite{Vla80a,me80,me82a,me84} and below).

Now we turn to the $\Rm_U$ --- to the inverse of
the $\Ru$-operation. It reads
\beq \Rm_U(\G)=\sum_{S\in \Wu\{\G\}}\Dm_U(S).\EQN{ms2b}\eeq
As a consequence of \re{ms2a}
we get that the $\Dm_U$-operation
also possesses the minimality property, i.e.
\beq \Du^{-1}\!<\g>=\sum_{i}Z_{i}^{(-1)}(<\g>,\ep)
{\bf  e}_{i}({\bf  q}^{\g},{\bf  m}^{\g}).\nnb \eeq
with
$K\,Z^{(-1)}_{i}=Z^{(-1)}_i$.

It was stressed by the authors of
refs. \ocite{T86,T88} that in the MS-scheme
$\Rm_U$-operation plays
a role which, in  a sense, is
similar to that of the Zimmermann identities
in the momentum subtraction scheme\foot{See in this connection the next
subsection.}.  It could be effectively employed
to express an unrenormalized
(or only partially renormalized) FI
as a sum of appropriately constructed MS-renormalized FI's.

The main idea of these applications of the $\Rm_U$-operation
is very simple and could be  illustrated best
by the following example.
Let the product $R_{U}\Rm_U(\G)$ act on an unrenormalized FI
$<\G>$. Simple manipulations based
on \re{r4} give  the identity
\beq <\G>=\sum_{S\in \Wu\{\G\}}
\left\{
\prod_{\g\in S }
\Dm(\g)<\g>\right\}\ast\Ru\!<\G/S>,\EQN{ms3}\eeq
which expresses the initial FI $<\G>$
in the form of a linear
combination of completely MS-renormalized FI's
times some constants  (divergent in the limit of
$\ep$ going to zero).

\def\i{{\bf i}}
\def\lp{'{}\!}

\sect{BPHZ renormalization and the $\Rm$-operation}
In this section we shall clarify the relation
between the  $\Rm$-operation and the good  old
Bogoliubov-Parasiuk renormalization prescription
which is based on the momentum
subtractions. It will be shown that the use of the
combinatorial technique above  developed uncovers a remarkable
hidden structure of this renormalization scheme.
We shall then employ this structure to obtain simple
regular derivations of  the Zimmermann identities and
the conversion formulas establishing a connection
between the minimal subtraction and momentum
subtraction schemes.

We begin, as usual, by fixing our notation. The BP
$R$-operation,  ${\cal  R}(\G)$, is defined in terms of
the corresponding $c$-operation $\de$ as follows
\beq {\cal  R}(\G)=\sum_{S\in \Wu\{\G\}}\de(S), \EQN{bp1} \eeq
\beq \de(\g)=-M(\g)\lp\,{\cal  R}(\g),\EQN{bp2}\eeq
where $M(\g)$ is a subtraction operator
(in fact another $c$-operation)
which associates with a given FI
\ $<\g>$ \ its truncated Taylor expansion in the external momenta
${\bf  q}^{\g}$~.
The  formal definition of $M$
is (note that this definition assumes that one
first chooses an initial FI $<\G>$ so that Feynman subintegrals
$<\g'>$, $\g'\in\G$ and, thereby, $\omega(\g')$ are well-defined)
\ice{
\begin{align*}
 u(x) &=
  \begin{cases}
   0        & \text{if } x \geq 0 \\
   1        & \text{otherwise}
  \end{cases}
  \\
 \tau &=
  \begin{cases}
   a+b+c+d  & \text{if } y < 0 \\
   1        & \text{otherwise}
  \end{cases}
\end{align*}
}
\beq
<\g> =
\begin{cases} 
\T^{\omega(\g')}_{{\bf  q}^{\g}}
<\g>, &\text{ if }  \g=\g'/S ,\,\g'\subseteq\G, \,<\g',\,S\in \Wu\{\G\},
\\
0, & \text{ if } \g  \text{ is an isolated vertex},
\end{cases}
\EQN{bp3}
\eeq

\beq 
\T^n_{\dots}=\sum_{i\geq 0}^{(n)}\T^{(i)}_{\dots},
\EQN{bp4}
\eeq
where
\foot{Throughout this subsection  it is assumed
that  Feynman integrals
encountered contain no massless lines and
thus are free of any IR
divergences.This qualification  ensures that
the operator $\T^{(i)}_{{\bf  q}^{\g}}$ is well
defined for any integer $i\ge0$.}
we have denoted by $\T^{(i)}_{\dots}$ the operator that
picks out the terms of the i-th order of the Taylor expansion
with respect to the corresponding variables, e.g.
\beq \T^{(n)}_\rho f(\rho)={1\over{n!}}\left({d\over{d\kappa}}
\right)^n
f(\kappa\rho)
\vert_{\kappa=0}
\nnb
\eeq
and
$\dsp\T^{(n)}_\rho f(\rho)\equiv0$ if $n<0$.
The complicated recurrence structure of the ${\cal  R}$-operation can be
explicitly resolved via the  forest formula \ocite{ZS65,Z69}
which has the form
\beq {\cal  R}(\G)=\sum_{F\in F \{\G\}}(-)^{\vert F \vert}M(F).\EQN{bp5}\eeq
The BPH theorem ensures that for every FI $<\!\G\!>$
without massless
lines the combination ${\cal  R}<\G>$ is finite
at $\ep\rightarrow 0$.

In view of \re{rm7}, it is clear that the inverse
${\cal  R}^{-1}$-operation assumes a remarkably simple
{\it explicit} form, viz.
\beq {\cal  R}^{-1}(\G)=\sum_{S\in \Wu\{\G\}}M(S).\EQN{bp6}\eeq
  
Below we shall see that it is
the simplicity of the ${\cal  R}^{-1}$-
operation that distinguishes the
BHZP renormalization prescription.

Indeed, as is  well known,
the concepts of oversubtractions and the so-called Zimmermann
identities are of vital importance for
 most applications of the BPHZ approach. Now we employ the general
formalism of the $R^{-1}$-operation to give a simple and regular way
of deriving  these identities and their generalizations.

Suppose we are given a rule which associates an integer $a(\g)$
with every 1PI subgraph $\g\subset\G$. Let us put, in addition,
$a(\g)=a(\g')$ for any graph $\g'=\g/S$
with $\g'\subseteq\G$ and $S\in \Wu\{\G\}$.
The ${\cal  R}$-operation with
oversubtractions corresponding to
the rule $a(\g)$ reads \ocite{Z69}
\beq {\cal  R}_a(\G)=\sum_{F\in \Fu\{\G\}}(-)^{\vert F\vert}
M_a(F),\EQN{bp7}\eeq
where the c-operation $M_a$ is defined as
\beq 
M_a<\g>=
\begin{cases}\T^{a(\g)}_{{\bf  q}^{\g}}<\g> & \text{if}
\g  \text{ is a  non-trivial 1PI graph from } {\cal  P}\{\G\}, 
\\
0, & \text{otherwise}. \phantom{aaaaa} 
\end{cases}
\EQN{bp8}
\eeq
The following proposition is valid \ocite{Z69}.
\\
{\bf Theorem} 3\ %
{\it
The Feynman integral
${\cal  R}_a<\G>$
is finite at $\ep=0$ if the
subtraction degrees $a(\g)$ satisfy the inequalities
\beq a(\g)\ge\omega(\g)+\sum_{\de\in S}(a(\de)-\omega(\de))\EQN{bp9}\eeq
for any 1PI subgraph $\g\in\G$ and any spinney $S\in \Wu\{\g\}$.}

An example of choosing the function $f(\g)$
meeting the constraint \re{bp9} is, in fact, provided the
${\cal  R}$-operation.

As  a direct consequence of the theorem, we get that
the $R$-operation ${\cal  R}_a$ makes also finite
every FI $\displaystyle{\g=
\left(\prod_{\g''\in S} P_{\g''}({\bf  q}^{\g''})\right)*<\g'/S }>$
with $\g'\subset\G$ and $\g>S\in \Wu\{\g\}$  {\it
provided} the degree of every polynomial
$P_{\g''}$ is less than or equal to $a(\g'')$. In what
follows we shall refer to such FI as {\it $a$-admissible} ones.

An equivalent form of \re{bp7} is
(cf. eqs.\re{bp5} -\re{bp6} )
\beq {\cal  R}_a(\G)=\left({\cal  R}_a^{-1}(\G)\right)^{-1},\EQN{bp10}\eeq
with
\beq {\cal  R}_a^{-1}(\G)=\sum_{S\in \Wu\{\G\}}M_a(S).\EQN{bp10p}\eeq

Let us now consider two different ${\cal  R}$-operations with
oversubtractions, viz. ${\cal  R}_a$ and ${\cal  R}_b$ such that
both functions
$a(\g)$ and $b(\g)$ are constrained by \re{bp9} and
the inequality
\beq a(\g)\ge b(\g)\EQN{bp11}
\eeq
holds for every $\g\subseteq\G$. At the level
of individual FI's
the Zimmermann identity may be considered as a conversion
formula to
express an ${\cal  R}_a$-normalized FI in terms of ${\cal  R}_b$-normalized FI's .
To derive the conversion relation it suffices to
elaborate a trivial
identity
\beq 
{\cal  R}_a(\G)\equiv{\cal  R}_b(\G){\cal  R}_b^{(-1)}(\G){\cal  R}_a(\G)\EQN{bp12}
\eeq
with the help of \re{r3} and \re{bp10p}. As a result we get
\begin{subequations}\label{3.32}
\beq {\cal  R}_a(\G)={\cal  R}_b(\G){\cal  R}_{b,a}(\G), 
\eeq
\beq {\cal  R}_{b,a}(\G)={\cal  R}_b^{(-1)}(\G){\cal  R}_a(\G)=
\sum_{S\in \Wu\{\G\}}\de_{b,a}(S), 
\eeq
\beq \de_{b,a}(\g)=(M_b-M_a){}'{\cal  R}_a(\g). 
\eeq
\end{subequations}
Here ${\cal  R}_{b,a}$ is an  $R$-operation,
with
\beq \de_{b,a}<\g>=\sum_{{i}_{\g}}\xi^{b,a}_{i_{\g}}{\bf  e}_{{i}_{\g}}({\bf  q}^{\g})
\EQN{bp13}
\eeq
and the coefficients $ \xi^{b,a}_{i_{\g}}$
being  finite at $\ep\rightarrow 0$
for every $a$-admissible FI $<\g>$ .
Indeed, it follows
from \re{3.32} that
\beq \de_{b,a}<\g>=\sum_{i>a(\g)}^{b(\g)}
\T_{{\bf  q}^{\g}}^{(i)}{\cal  R}_a<\g>,\EQN{bp14}
\eeq
with ${\cal  R}_a<\g>$ being finite due to Theorem 3.
We now
combine \re{3.32} and  \re{bp14} to come to the
following conversion formula \beq {\cal  R}_a<\G>=\sum_{S\in
\Wu\{\G\}}\de_{b,a}<S>*{\cal  R}_b<\G/S>\EQN{bp15}\eeq
or, explicitly,
\beq {\cal  R}_a<\G>=\sum_{S\in \Wu\{\G\}}\sum_{\i}\prod_{\g\in
S}\left(\xi^{b,a} _{i_{\g}}\right)
{\cal  R}_b<\G/S>^{\i}.\EQN{bp16}\eeq

Let us turn to the derivation of
the conversion formulas which allow
one to express easily a minimally subtracted FI in the form of
an appropriate linear combination of FI's renormalized by
the ${\cal  R}_a$-operation, and vice versa. Let us define
\begin{subequations}\label{3.37}
\beq R_{a,U}(\G)={\cal  R}_a^{-1}(\G)\Ru(\G)\label{3.37a} 
\eeq
\beq R_{U,a}(\G)=\Ru^{-1}(\G){\cal  R}_a(\G).\label{3.37b} 
\eeq
\end{subequations}
An almost literal repetition of the above argument
shows that the corresponding c-operation  is
\begin{subequations}\label{3.38}
\beq \delta_{a,U}<\g>=M_a {}'\Ru<\g>+\Du<\g> \label{3.38a} \eeq
while
\beq \delta_{U,a}<\g>\equiv\delta^{-1}_{a,U}<\g>=
\delta_{a,U}(\g)\sum_{F\in \Fu\{\g\}}(-1)^{\vert F\vert}
\delta_{a,U }(F)<\g>.\label{3.38b}  \eeq
\end{subequations}
Thus,
\begin{subequations}\label{3.39}
\beq \Ru<\G>=\sum_{S\in
\Wu\{\G\}}\de_{a,U}<S>*{\cal  R}_a<\G/S>,\label{3.39a} \eeq
and
\beq {\cal  R}_a<\G>=\sum_{S\in %
\Wu\{\G\}}\de_{U,a}<S>*\Ru<\G/S>.\label{3.39b} \eeq
\end{subequations}

As a consequence of Theorem 2 the
R-operations $R_{a,U}$ and $R_{U,a}$ are
evidently finite\foot{That is  both c-operations
 $\delta_{a,U}(\g)$ and  $\delta_{U,a}(\g)$ transform any
$a$-admissible FI $<\g>$ into  polynomials
in ${\bf  q}^{\g}$ finite at $\ep\to0$.}
on the class of $a$-admissible FI's if and only
 if the subtraction degrees $a(\g)$ meet condition
 \re{bp9}.  It is worth noting that Theorem 3 follows
directly from this observation and the conversion
relation \re{3.39b}.

\chap{$R^*$-OPERATION}
The purpose of this section is
twofold. First, we formulate an extended version of the
$R^*$-operation which minimally removes both the UV and IR
divergences from a  (dimensionally regularized) euclidean
Feynman integral with arbitrary set of external momenta
(including the exceptional ones) and  formally expanded
in some of its  masses and external momenta. Second,
we employ the forest technique developed above to simplify
considerably the apparatus of the $R^*$-operation by
explicitly expressing the corresponding IR counterterm
operation in terms of its UV counterpart. This extended formulation
of the $R^*$-operation is useful for the studying asymptotic
expansions of generic Feynman integrals (see the next
section) and computing the UV renormalization constants (see
below).

Since the $R^*$-operation is  comparatively a newcomer in the
renormalization market we begin with a special introductory
section which provides the reader with a short overview of
the main ideas of the $R^*$-operation and a simple illustrative
example of its use in calculations.
\sect{$R^*$ - primer}
A  lot  of information   about    the     structure
of the renormalized quantum field theory is contained in the
renormalization    group   (RG)    functions
(the $\beta$-functions and  anomalous  dimensions).
These  functions, in their turn,  are  expressible    through
some     combinations    of     UV renormalization
constants (see  any   textbook  on
QFT).  Historically, the $R^*$-operation was
invented \ocite{me82a} in the attempt to get rid of some
limitations of the so-called  IR rearrangement  (IRR)
trick \ocite{Vla80a}, where, in
fact, it was first demonstrated that   the   problem
of   evaluating the UV renormalization   constants can  be
drastically   simplified by proper use of  the
rich    possibilities    offered  by   the
MS-scheme. Coupled  with the  $R^*$-operation, the  IRR trick
became  a  regular  powerful  method
of computing the  UV renormalization
constants \ocite{me82a,me84}.

The starting ideas both of the IRR trick
and of the $R^*$-operation
can be illustrated best exposed by
considering a couple of simple
examples.  Let us begin with a qudratically divergent FI,
\beq <\G_1>(q,m)=\int {{d^\mu k_1d^\mu
k_2}\over{((q-k_2)^2+m^2) ((k_1-k_2)^2+m^2)
({k_1}^2+m^2)}}\EQN{411}
\eeq
The corresponding UV counterterm reads
\beq \Du<\G_1>=Z_m m^2+ Z_2 q^2\EQN{412}\eeq
where $Z_m$ and
$Z_2$ contribute to the mass and wave functions
renormalizations, respectively, in the
$\varphi^4$-model\foot{Throughout this subsection we work in
four-dimensional space-time and hence set $D_0=4$.}.  After
differentiating $<\G_1>$  with respect to $m^2$, one comes to
another, now only logarithmically divergent, FI:
\beq <\G_2>(q,m)=\int {{d^\mu k_1d^\mu
k_2}\over{((q-k_2)^2+m^2) ((k_1-k_2)^2+m^2)
({k_1}^2+m^2)^2}}.\EQN{413}\nnb \eeq
  
The differentiated version of
eq.\re{412} is
\beq \Du<\G_2>=-Z_m /3.\EQN{414}\eeq
The idea of the IRR trick is quite simple. Since the UV
renormalization constant of a logarithmically divergent FI
does not depend on any dimensional parameters and since our
aim is to calculate just this constant, nothing prevents us
 from introducing auxiliary external momenta and/or masses
along with setting zero the initial mass and external
momentum $provided$ that this does not lead to any IR
divergences. In many cases,  the resulting FI becomes
simpler for calculation. For the FI in question this can be
easily achieved if we introduce a new non-zero external
momentum $q'$ to flow through {\it one} of the two lines in
the series ( both with one and the same propagator
$1/(k_1^2+m^2)$ in \re{413}) and set $q$ and
$m$ zero. Indeed, the resulting  FI 
\beq <\G_3>(q')=\int {{d^\mu
k_1d^\mu k_2} \over{k_2^2 (k_1-k_2)^2(k_1-q')^2
k_1^2}}\EQN{414a}\eeq
is readily computed in terms of
$\G$-functions (see, e.g. ref. \ocite{me80}).

Unfortunately, the condition of that the IR
divergences do not appear restricts considerably the range
of applicability of the IRR trick, since for complicated
FI's this prevents one from transforming a given FI to the
simplest form (see, e.g. ref. \ocite{me82a}).

A radical generalization of IRR, free of any limitations of
this kind, is based on the $R^*$-operation.  To
understand the essence of this approach, let us  try  to
compute $Z_m$ starting  directly from FI $<\G_2>(q,m)$ and
this time $avoiding$ any manipulations with external
momenta. By definition of  $\D_U$,  we have
\beq \D_U<\G_2>=K_\ep \Bigl[<\G_2>(q,m)+ Z_\g \int {{d_\mu k_1}
\over {(k_1^2+m^2)^2}}\Bigr],\EQN{416}
\eeq
where  $Z_\g=-(16\pi^2\ep)^{-1}$ and the second term subtracts
the UV subdivergence corresponding to the subintegration
over the loop momentum $k_2$. If now we tried to set
$m=0$ in the corresponding integrand, then the integration
over the  small $k_1$ region would lead to an IR divergence.
Within the DR this IR divergence shows
up as an extra pole in $\ep$.  As a result, a
straightforward calculation of the rhs of \re{416} would
give the wrong result, containing  spurious IR poles  along
 with the true UV ones.  Now there arises a natural
question: can one extract the IR pole and remove
it by  subtracting from $<\G_2>(q,0)$ an IR counterterm
local in the $x$-space and  not in the $p$-space?  The
affirmative answer was given in
refs. \ocite{me82a,me84} where the  $\Rt$-operation with
the necessary properties was constructed.  It may be easily
checked that the combination
\beq f(k_1)=k_1^{-4}+\widetilde Z\delta(k_1),\ \
\hbox{with} \ \ \ \widetilde Z=(16\pi^2\ep)^{-1}\EQN{417}\eeq
does not
lead to any IR poles after a formal D-dimensional integration
with a function  smooth at $k_1=0$.
This means that the
pole part of the integral
\beq {}'R\Rt<\G_2(q,0)>= \int \Bigl[\int
(q-k_2)^{-2}(k_1-k_2)^{-2}f(k_1) d_\mu k_2 +Z_\g
f(k_1)\Bigr]d_\mu k_1\EQN{418}\eeq
is determined by the behavior of
the integrand in the region of large $k_1,k_2$ and coincides
with the UV counterterm we are looking for.  In \re{418}
the symbol $\Rt$ stand for the IR $\Rt$-operation that
removes the IR divergencies by x-local counterterms and
in our case comes to replacing $k_1^{(-4)}$ by
combination \re{417}.

Unfortunately, the representation of the IR counterterms
through $\delta$-functions of some loop momenta becomes rather
cumbersome and unnecessarily complicated in general case.
Another approach that is much more convenient and powerful
was first outlined in ref. \ocite{me87b} and will be presented
below.

\sect{Generalized forest technique}
In this subsection we extend some of our previous
considerations to the more general case of spinneys and
forests formed by arbitrary connected graphs.  Though
such a generalization is quite  straightforward it
proves to be quite useful in the treatment of the
$R^*$- operation and asymptotic expansions of Feynman
integrals.

Let us denote as ${C}\{\G\}$ the collection of
all connected subgraphs of a graph $\G$.
From now on
a spinney (forest) of $\G$ will refer to
a collection of connected
non-intersecting (non-overlapping) non-empty
subgraphs of $\G$. Thus, if $S$ is a member of $W\{\G\}$
then for any  pair $\g,\g^\prime$,$\g\not=\g'$
from $S$ one has
\beq 
\g\in C, \ \g'\in C\{\G\}\ \hbox{and}\ \ %
\g\cap\g'=\G_\emptyset.\EQN{gf1}
\eeq
It is an easy exercise to see
that all the reasoning of sections
3.1 and 3.2  can be applied without any modifications to
these extended definitions. Note also that even for  a
one-particle-reducible  graph
$\g$  one has $\D(\g)\not\equiv 0$,  in spite of
the fact that by definition
$\D<\g>\equiv0$. The reason is
that for a subgraph $\delta\subseteq\g$ such that
the reduced graph $\g/\delta$ is 1PI
$\Delta(\g)<\g/\delta>$ is $\D<\g/\delta>$
and, hence, can
be non-zero.

Let $C'\{\G\}\subseteq C\{\G\}$ be an arbitrary subset
of the family of connected subgraphs of the graph $\G$ and
$W'\{\G\}\subseteq W \{\G\}$
$\Bigl(F'\{\G\}\subseteq F\{\G\}\Bigr)$
stand for the collection of all
spinneys (forests)
with their elements from $ C'\{\G\}$.
Let us define
\beq R (W'\{\G\},\D)=
\sum_{S\in W'\{\G\}} \D(S),\EQN{gf3}\eeq
\beq R (F'\{\G\},\D)=
\sum_{F\in F'\{\G\}} \D(F),\EQN{gf4}\eeq
where $\D$ is a $c$-operation.
A literal repetition of the argument given in
section 3.3 immediately shows that
\beq \Rm(W'\{\G\},\D)=
R(F'(\{\G\},-\D),\EQN{gf5}\eeq
\beq \Rm(F'\{\G\},\D)=
R(W'(\{\G\},-\D).\EQN{gf6}
\eeq

For any spinney $S\in W\{\G\}$ we  denote
\beq  C'\equiv C^\tld\{\G,S\}
=\{\g\vert S\sim\g\in C\{\G\}\},\nnb \eeq
\beq  C''\equiv C^\succ\{\G,S\}
=\{\g\vert S\prec\g\in C\{\G\}\},\nnb \eeq
\beq  C''{}'\equiv C^{\gs}\{\G,S\}
=\{\g\vert S\ls\g\in C\{\G\}\},\nnb \eeq
while

\endgraf\noindent
\centerline {
$W^\tld\{\G,S\}$,\ \
$W^\succ\{\G,S\}$,\ \
$W^{\gs}\{\G,S\}$,\  \
$F^\tld\{\G,S\}$,\   \
$F^\succ\{\G,S\}$\ \hbox{and}   \
$F^{\gs}\{\G,S\}$ }\     \
\endgraf\noindent
will stand for
the respective families
\vspace{4mm}

\centerline {
$W'\{\G,S\}$,\ \
$W''\{\G,S\}$,\ \
$W'''\{\G,S\}$,\ \
${\cal  F}'\{\G,S\}$,\   \
${\cal  F}''\{\G,S\}$ \ \hbox{and}  \
${\cal  F}'''\{\G,S\}$.
}
\vspace{4mm}
\noindent
{\bf Theorem 4.}
{\it For an arbitrary spinney $S\in W\{\G\}$ the following
relations are true
\beq R(W^{\gs}\{\G,S\},\D)<\G/S>=
R(W\{\G/S\},\D)<\G/S>,\EQN{gf7}\eeq
\beq R(W^{\gs}\{\G,S\},\D)=
R(W^\tld\{\G,S\},\D) R(W^\succ\{\G,S\},\D),\EQN{gf8}\eeq
\beq \Rm(W^{\gs}\{\G,S\},\D)= R(F^\succ\{\G,S\},-\D)
R(F^\tld\{\G,S\},-\D).\EQN{gf9}\eeq
  }

\noindent
{\bf Proof.}  Relation \re{gf7} is nothing but a
shorthand of \re{r4}. Eq.\re{gf8} is a direct
consequence of the fact that every spinney $S$ from
$W^{\gs}\{\G,S\}$ can be unambiguously presented in a form
\beq
S=S^{\succ}\cup S^{\tld}\ \ \mbox{with} \ \
S^\succ\in W^\succ\{\G,S\} \ \  \mbox{and} \ \
S^\tld\in W^\tld\{\G,S\}
\nnb
\eeq
and of the observation
that no graph $\g$
from $ C^\tld\{\G,S\}$ can have a proper
subgraph $\g'\subset\g$ such that $\g'\in C^\succ\{\G,S\}$.
In  order to prove \re{gf9} it is sufficient to make
use of \re{gf8} and \re{gf5}.

\sect{Extra notation and definitions  for Feynman integrals}

Let $<\G>({\bf  q},{\bf  m})$ be a dimensionally regularized FI corresponding to a
(connected) graph $\G$.  Its external momenta may  obey some
extra linear constraints of the form
\beq 
\sum_{l\in {\rm c}_i}\varphi_{\G}(l)q_l=0,\ \ \ \ \ \ \ \ \ \ \
\ \ \ {\rm c}_i\in{\bf  c}.  \EQN{en1}
\eeq
Here  ${\bf  c}=\{{\rm c}_i\}$ is a
collection of subsets of ${\cal  E}_\G$ which is to comprise the
empty set and  ${\cal  E}_\G$; this latter constraint is due to the
total momentum conservation.  Without essential loss of
generality, we shall assume that for any pair $i,i'$ such
that ${\rm c}_i\cap {\rm c}_{i'}=\emptyset$\ \ or    \ \ ${\rm c}_i\subset
{\rm  c}_{i'}$, one has either $({\rm c}_i\cup c_{i'})\in{\bf  c}$ or $({\rm c}_i'\b
c_{i})\in{\bf  c}$, respectively.  The case of non-exceptional
external momenta corresponds to the choice
${\bf  c}=\{{\cal  E},\emptyset\}$.

If $\g$ is a connected graph from ${\cal  P}\{\G\}$, then the
relevant set of external momenta ${\bf  q}^{\g}$ will be subjected
to linear constraints induced by  \re{en1} and the total
momentum conservation i.e. 
\beq \sum_{l\in
{\rm c}_i}\varphi_{\g}(l)q^{\g}_l=0, \ \ \ \ \ \ \ \ \ \ \ \ \ \ \
{\rm c}_i\in{\bf  c}^{\g} \EQN{en2}
{},
\eeq
where
\beq {\bf  c}^{\g}=\{{\rm c}_i\vert
{\rm c}_i\subseteq{\cal  E}_{\g},{\rm c}_i\in{\bf  c}\} \cup
\{{\cal  E}_{\g}\b {\rm c}_i\vert
{\rm c}_i\subseteq{\cal  E}_{\g},{\rm c}_i\in{\bf  c}\}.\EQN{en3}\eeq
Note that even if the external momenta
$q_l\in{\bf  q}$ are non-exceptional
the vectors from ${\bf  q}^{\g}$ will, in general, form  an
exceptional set of external momenta. Indeed, if, for instance,
${\cal  E}\subseteq{\cal  E}_{\g}$,
then
${\bf  c}^{\g}=\{{\cal  E},{\cal  E}_{\g}\b{\cal  E},\emptyset\}$ for
${\bf  c}=\{{\cal  E},\emptyset\}$, as it follows from \re{en3}.

A subset ${\bf  q}'\subseteq{\bf  q}^{\g}=
\{q_l\vert l\in{\cal  E}'\subseteq{\cal  E}_{\g}\}$
is said to be a {\it right} ({\it r}-)subset of
${\bf  q}^{\g}$ if the decomposition
${\bf  q}^{\g}={\bf  q}'\cup({\bf  q}^{\g}\b{\bf  q}')$ is well-defined
with respect to constraints \re{en2}. In other words,
each of the constraints should remain valid after
all the momenta from ${\bf  q}'$ are multiplied by an auxiliary
parameter.  The requirement is equivalent to demanding that
the intersection $({\cal  E}'\cap {\rm c}_i)\subseteq{\bf  c}^{\g}$ for every
${\rm c}_i$ from ${\bf  c}^{\g}$.

Suppose  we are given a partition of
external momenta and masses of the FI $<\G>$ of the 
form\foot{It is convenient to assume that all the massless lines from
${\cal  L}$ belong to ${\cal  L}_0$.}
\beq {\bf  q}={\bf  q}_1\cup{\bf  q}_0\ ,\  \ \ \ \ \ \ \ \
{\bf  m}={\bf  m}_1\cup{\bf  m}_0,\EQN{en4}
\eeq
\beq {\bf  q}_i=\{q_l\vert l\in{\cal  E}^i\},\ \ {\bf  m}_i=\{m_l\vert l\in{\cal  L}^i\},\ \
{\cal  E}^0\cap{\cal  E}^1=\emptyset
={\cal  L}^0\cap{\cal  L}^1 ,\ \ i=0,1,\nnb 
\eeq
with ${\bf  q}_1$ and ${\bf  q}_0$ being {\rm r}- subsets of ${\bf  q}$.
For every connected graph $\g\in{\cal  P}\{\G\}$ we define
\beq {\bf  q}_i^{\g}=\{q_l^{\g}\vert l\in{\cal  E}^i_{\g}\}
\ \ \hbox{\rm{and}} \ \ {\bf  m}_i^{\g}=\{m_l\vert l\in{\cal  L}^i_{\g}\},
\ \ i=0,1,\nnb 
\eeq
where
\beq {\cal  E}_{\g}^1={\cal  E}^1\cap{\cal  E}_{\g},\ \ {\cal  E}_{\g}^0={\cal  E}_{\g}\b{\cal  E}^1_{\g}
\ \ \hbox{\rm {and}} \ \ {\cal  L}^i_{\g}={\cal  L}^i\cap{\cal  L}_{\g},
 \  \ \ i=0,1. \EQN{en5}
\eeq

Let $\g$ be a connected graph from ${\cal  P}\{\G\}$.
If the external momenta ${\bf  q}^{\g}_1$ form an {\rm r}-subset
of ${\bf  q}^{\g}$, then $\g$ will be termed as a 1-{\it right} graph.
A vertex $v\in{\cal  V}_{\g}$ is said to be $1$-$hard$, if
\beq \sum_{l\in{\cal  E}^{1}\cap{\cal  E}(\dot v)}q_l\not\equiv 0\nnb \eeq
and/or at least one massive line from
${\cal  L}^1_{\g}$ has $v$ as its incident vertex;
otherwise $v$ will be termed $1$-$soft$. For example,
if ${\cal  L}^0={\cal  E}^0=\emptyset$, then
for a given vertex of $\G$ to be $1$-hard, the algebraic sum
of external
momenta
flowing into it should be non-zero,
or the vertex should be incident at least to one massive
internal line.  If $\g$ is 1-right, then it will
be referred to as
\endgraf \noindent
(i) 1-soft, if  ${\cal  V}_\g$ comprises
no $1$-hard vertices;
\endgraf
and
\endgraf
\noindent (ii)
{\it $1$-irreducible}, if there is no such
line $l, \ {\cal  L}^1\not\ni l\in\Lg$,
that the graph $\g-l$ obtained by deleting $l$ from $\g$
remains $1$-right and $c(\g-l)>c(\g)$. In other words
for $\g$ to be $1$-irreducible it must not involve a cut-
line $l$ whose corresponding internal momenta $p_l^\g$ can
 be identically expressed (via linear constraints
 on ${\bf  q}^{\g}$) exclusively  in  terms  of  the momenta   from   the
set   ${\bf  q}_0^{\g}$.  Note that a
$1$-soft graph $\g$ could be $1$-irreducible if and only if
it is 1PI. Finally, a spinney $S\in W\{\g\}$
consisting of 1-right subgraphs of $\g$
will called \endgraf\noindent (i) a 1-$uniting$ one, if every
1-hard vertex from ${\cal  V}_{\g}$ belongs to some graph from $S$;
\endgraf \noindent
and
\endgraf \noindent
(ii) a 1-$hard$ one, if it comprises
no 1-soft graphs.

A very useful transformation of a Feynman integral is its
expansion in a  (formal) Taylor series with respect to
some of its external momenta and masses. Suppose one is
going to expand the FI $<\G>({\bf  q},{\bf  m})$ in masses and momenta
from the collections ${\bf  q}_0$ and ${\bf  m}_0$, respectively. The
operator $t_0$ which performs the procedure is defined as
\beq t_0=\sum_{n\ge
0}^{\infty}{\xi}^{n}t_0^{(n)}, \ \ \
t_0^{(n)}=\T^{(n)}_{{\bf  q}_0,{\bf  m}_0}.\EQN{en6}
\eeq
Thus
\beq t_0^{(n)}<\G>={1\over{n!}}{\Bigl({d\over{d\xi}}\Bigr)}^n
<\G>({\bf  q}_1\cup\xi{\bf  q}_0,{\bf  m}_1\cup\xi{\bf
  m}_0)\vert_{\xi=0},\EQN{en7}
\eeq
and
\beq t_0^{(n)}<\G>=\sum_i
<\G>_i({\bf  q}_1,{\bf  m}_1){\bf  e}_i({\bf  q}_0,{\bf  m}_0),
\EQN{en8}
\eeq
with ${\bf  e}_i({\bf  q}_0,{\bf  m}_0)$ being basis vectors of a (mass)
dimension $n$ from $A({\bf  q},{\bf  m})$.
Here several  comments are in order.

\noindent
(i) The differentiation
with respect to $\xi$ in \re{en7} may be carried out in
three ways.  In first place, one could simply differentiate
the  FI, which is a smooth function of $\xi$ at $\xi\not=0$.
Another  way is to differentiate the corresponding
$integrand$. Here in turn two options arise. First, we might
employ the momentum space representation as described by
\re{221}. (It is this option, which is always used in
carrying out practical calculations because of flexibility
and compactness of this representation.) Second, it is
possible to use the $\alpha$-parametric representation.
In what follows we shall
follow to  the last option since this allows  for a
direct interpretation of $<\G>_i({\bf  q}_1,{\bf  m}_1)$ as a FI
corresponding to the $same$ graph $\G$ with the same
but somewhat differently treated $\alpha$-parametric
integrand\foot{Generally speaking, it is by
no means obvious that these ways are equivalent.
Fortunately, this is the case within the dimensional
regularization \ocite{me83c}.}.

\noindent
(ii) The operation of setting $\xi$ zero
is meant to be a formal one, i.e., it should act on the
differentiated integrand.
\endgraf \noindent
(iii)The infinite series in the rhs of \re{en6} is to
be understood as a formal one; so no question about its
convergence may arise.  Moreover, in every application where
 the operator $t_0$ can appear, the terms of too high orders
in $\xi$ may be dropped for one reason or another, while the
remaining (finite) series in $\xi$ is taken at $\xi=1$.
Having this in mind we put $\xi=1$ in all formulas below.

\sect{$R^*$-operation in the MS-scheme}
Given  an arbitrary $c$-operation, $\D_g$, we define
another $c$-operation, $\D_{\dot g}$, according to the
following rule
\beq
\D_{\dot g}<\g>=
\begin{cases}
\D_{g}<\g>&if \, {\cal  L}_\G\not=\emptyset;
\\
<\g>& \text{ if  } \g  \text{  is an isolated vertex.}
\end{cases}
\EQN{441}
\eeq

Further, let
$\widetilde{{C}}\{\G\}$
stand for a collection of graphs
 which could be produced
from a graph $\G$
by reducing some non-empty (possibly
disconnected) subgraphs of $\G$.
In other words,
\beq \Ct\{\G\}=\{\g\vert\g=\G/S, \ S\in W\{\G\}\}.\nnb \eeq
  
It is  worth noting that $\G\in\Ct\{\G\}$, since
by definition $\G/S_\emptyset=\G$,  and
if $\G$ is disconnected, then  for
every element  $\g\in\Ct\{\G\}$  the
number of its ${\rm c}$-component
will  be equal to that of $\G$.
Let us also define $\Wt\{\G\}$ as
a subset of $W\{\G\}$  comprising  all
(minimal) spinneys which generate
exactly the set $ \Ct\{\G\}$ as a result of the operation
$S\longrightarrow\G/S$.

The $\ct$-operation $\Dt_{{}_I}(\g)$ --- the
"infrared" counterpart
of $\D_{{}_U}$  --- is defined as
follows
\beq \Dt_{{}_I}(\g)\,t_0<\G>=
\begin{cases}
t_0\bigl(<S_{\g}>\bigr)\ast\D_{\dot I}<\G/S_{\g}>,%
& \text{ if } \g\in\Ct\{\G\} \text{ and }
\\ & S_{\g} \text{ is 1-uniting}
\\ 0,& \text{ otherwise}, 
\end{cases}
\EQN{442}
\eeq
where $S_{\g}$ is the (unique) spinney from
$\Wt\{\G\}$ such that $\G/S_{\g}=\g$.
Here  $\D_{{}_I}$  such is a $c$-operation that
for any collection of vertex polynomials
$\{ P_\g({\bf  q}_0)\vert \g\in  S_{\g}\}$ the combination
\beq -\bigl(\prod_{\g\in S_{\g}} P
_\g\bigr)*\D_{{}_I}<G/S_{\g}>\equiv
-\D_{{}_I}\Bigl(\bigl(\prod_{\g\in S_{\g}} P
_\g\bigr)*<G/S_{\g}>\Bigr)\nnb \eeq
  
is the overall  IR divergence of a
(tadpole\foot{This is the case since
$\g=\G/S_\g$ is 1-soft owing to the
definition \re{442}.}) FI:
\beq -t_0\Bigl(\bigl(\prod_{\g\in S_{\g}}
P_\g\bigr)*<G/S_{\g}>\Bigr)\nnb \eeq
  
(more details about the possible
choices of  $\D_{{}_I}$ are presented below).

 Another useful version of \re{442} reads
\beq 
\!\!\!\!\Dt_{{}_I}(\g)t_0^{(n)}<\G>=
\begin{cases}
\bigl(t_{0}^{(n')}<S_{\g}>\bigr)*\D_{\dot I}<\G/S_{\g}>,
& \text{ if } \g\in\Ct\{\G\}  \text{ and }  S_{\g}
\\ & \text{ is 1-uniting; }
\\ 0,& \text{otherwise, } 
\end{cases}
\EQN{443}\eeq
with\foot{It is understood here that in the evaluation
of the UV index of the FI $<\G/S_{\g}>$ the unit vertex
polynomial is associated  with
every vertex $v_\delta\in{\cal  V}(\G/S_{\g}),\ \delta\in S_{\g}$.}
$n'=n-\omega(\G/S_{\g})$.

The  $\D_{{}_I}$-operation
has the following easily checked properties:

\noindent
(i)$ \ \ \ \ \Dt_{{}_I}(\G)t_0<\G>= t_0<\G> \ \hbox {if} \ \ \ \G
\ \ \ \hbox {is a trivial graph;}$

\noindent
(ii)$  \ \ \ \Dt_{{}_I}(\G)t_0<\G>= \D_{{}_I}<\G> \ \hbox {if} \ \ \ \G
\ \ \ \hbox {is a 1-soft, non-trivial graph;}$

\noindent
(iii)  \ \ if the spinney $S_{I}$
comprises at least one $1$-reducible element, then

\noindent
\phantom{(iiiii) }$\Dt_{{}_I}(\g)t_0<\G>= 0$
due to the very FI $t_0<S_{\g}>$
vanishes.
\\
Finally, we  define the combined action of
$\D_{{}_U}$ and $\Dt_{{}_I}$ operations in the natural way:
\beq \D_{{}_U}(S)
\Dt_{I}(\g)\equiv
\Dt _{I}(\g)
\D_{{}_U}(S),\EQN{444}
\eeq

\beq \D_{{}_U}(S)\Dt _{I}(\g)t_0<\G>=
\left(t_{0}\D_{{}_U}(S)<S_\g>\right)*
\D_{\dot I}<\G/S_\g>\EQN{445}
{}.
\eeq
It follows from \re{445} that the product
$\D_{{}_U}(S)
\Dt _{I}(\g)$ can be non- zero if and only if
$S\leq S_\g $.
This  condition  reflects the simple fact that no propagator
can contribute to both UV and IR divergencies
simultaneously.

Now we are sufficiently equipped to introduce
the $R^*$-operation which subtracts   all kinds of divergences
from an (euclidean) dimensionally regularized FI formally
expanded in some of its external momenta and masses. We define
\beq R^*(\G)=R_{{}_U}(\G)\Rt_{{}_I} (\G)\equiv
\Rt_{{}_I}(\G)R_{{}_U}(\G)\EQN{446}
\eeq
where we have introduce
the  "infrared" $\Rt_{{}_I}$-operation which subtracts
the IR divergences only, and does not care for any UV ones:
\beq \Rt_{{}_I}(\G)=\sum_{\g\in \Ct\{\G\}}
\Dt_{{}_I}(\g ) \equiv\sum_{S\in \Wt\{\G\}}
\Dt_{{}_I}(\G/S) \EQN{447}
\eeq
Eqs.~(\ref{442},\ref{447}) imply that the
$R^*$-normalized FI $\ R^*t_0<\G>$
can be presented in the following convenient
form
\beq R^*t_0<\G>=\sum_{S\in \ {}_u\! W_1\{\G\}}
\ \sum_{S'\in W\{\G\}}^{ S'\le S}
\D_{{}_U}(S')\,t_0<S>*\D_{\dot I}<\G/S>
\EQN{448}
\eeq
\noindent or, equivalently,
\beq R^*t_0<\G>=\D_{\dot I}(\G)\sum_{S\in \ {}_u\! W_1\{\G\}}
\ \sum_{S'\in W\{\G\}}^{ S'\le S}
\D_{{}_U}(S')\,t_0<S>*<\G/S> \EQN{448b}
\eeq

Here
${}_u\! W_{1}\{\G\}$ stands for  the collection of
1-uniting spinneys from $\Wt\{\G\}$, which comprise
$1$-irreducible elements only.

Note that these  definitions coincide with
those of ref. \ocite{me84} in a
particular case where ${\bf  q}_0=\emptyset$,
${\bf  c}=\{{\cal  E},\emptyset\}$,
 and the set ${\cal  L}_0$
consists of massless lines only (this means that,
acting on the very
FI $<\G>$, the operator $t_0$ reduces to the unit one,
and that the external momenta are non-exceptional ones).
 We shall not develop the corresponding argument since
(\ref{447},\ref{448}) are much more general than definition
(8) of ref. \ocite{me84}  and should be considered in their
own right. Moreover, later we shall show that
eq.\re{448} provides one with a very convenient regular
algorithm to evaluate UV counterterms --- the problem which
in fact has generated the very idea of the $R^*$-operation.

{\bf Theorem 5.}\
{\it There exists such a choice of the $c$-operation
$\D_{{}_I}$ that the
$R^*$-operation defined by eq.\re{447} makes finite
arbitrary dimensionally regularized FI
$\ t_0<\G>$ at  $\ep \rightarrow 0$.
}

The  proof of this statement  will be presented in the next
section.  Now we use it to clarify the concept
of IR counterterms and to   construct  an algorithm for their
evaluation.

Let us choose ${\cal  E}^0={\cal  E}$ and ${\cal  L}^0={\cal  L}$. In this case
the operator $t_0$ converts the (now  1-soft!)
 FI $<\G>$ into a sum of massless tadpoles which are to
be put zero within the DR. To put it in
another way: in the absence of any dimensional parameters to
counterbalance the factor $\mu^{2\ep}$ (masses and external
momenta do not count due to our choice of $t_0$), the  UV and IR
divergences have no choice but to cancel each other.  Had we
subtracted the UV divergences with the help of the
$R_{{}_U}$-operation, this fine tuning would surely disappear
 --- the FI $R_{{}_U} t_0<\G>$ is free of any UV
singularities but still suffers from the IR ones,  and, thus
is, in general, not finite at $\ep\to0$. On
the other hand, Theorem 5 states that $R^* t_0<\G>$ must
be well-defined in this limit. It is thus only natural to fix
unambiguously the IR
$\D_{{}_I}$-operation by demanding

\beq R^*t_0<\G>=0\EQN{449}\eeq

for every 1-soft FI $<\G>$. It will be shown soon that this
normalization condition leads to great simplifications in all
applications of $R^*$-operation. To begin,  we shall
demonstrate  that  this convention immediately leads to  an
explicit expression for the IR $\D_{{}_I}$-operation in terms
of its UV counterpart.

Indeed, if $\G$ is
a 1-soft graph then each of its connected
subgraphs is 1-soft, too.
This means, in particular, that the combination
$t_0\bigl(\Du(S')<S>\bigr)$ is non-zero if
and only if $S'=S$. This observation allows one
to use \re{448} together with  definition
\re{441} in order to rewrite the normalization condition
\re{449} in the form
\newcommand{\lGr}{\left(\Gamma\right)}
\beq 
\Du\lGr<\G>+\Di \lGr {}'\Ru<\G>\equiv 0.
\EQN{4410}
\eeq
The equation holds for $any$  FI\  $<\G>$,
whence it follows that
\beq \Di(\G)=-\Du(\G)\Rm_{{}_U}(\G),\EQN{4411}\eeq
or, equivalently,
\beq \Di=\Du^{-1}.\EQN{4412}\eeq
  
Now we are going to derive a more compact representation
for the action of the $R^*$-operation upon $t_0<\G>$.  We write
\beq R^*t_0<\G> \nnb \eeq
\bea
&=&\sum_{\ \ \ \ S\in \
{}_u\! W_{1}\{\G\}} \left(\Ru t_0<S>\right)*\D_{\dot I}<\G/S>
\\
&=& \Ru t_0<\G>+\sum^{S \not=\G}_{S\in   \  {}_u\! W_{1}\{\G\}}
\left(\Ru t_0<S>\right)
  *\Di<\G/S>.
\EQN{4413}
\eea

Every spinney $S\in \  {}_u\!W_{1}\{\G\}$ can be unambiguously
represented in the form
\beq
S=S'\bigcup S'',\ \ \ \text{ where } \ \ \ S'\sim S'',
S'\in \  {}_u\! W^h_{1}\{\G\} \ \text{ and  }\ S''\in W^s_{1} \{\G\}.
\nnb \eeq
\noindent
Here
$W_{1}^h\{\G\}\left(W_{1}^s\{\G\}\right)$ stands for the
collection of all spinneys from $\Wt\{\G\}$ that include
only $1$-hard(soft) elements, while  ${}_u\! W_{1}^h\{\G\}$ is
composed of $1$-uniting spinneys from
$W_{1}^h\{\G\}$.  As a result, \re{4413}
could be presented as follows
\begin{align}
R^*t_0<\G>&=\Ru t_0<\G>+ 
\\
\sum^{S'\not=\G}_{S'\in  \ {}_u\!W^h_{1}\{\G\}} %
\Ru t_0<S'>&*\sum^{S''\sim S'}_{
 \ \ \ \  S''\in W^s_{1} %
\{\G\}}\Di(\G)\Du(S'')<\G/S'>
\\ 
 &=\Ru t_0<\G>+ \cr\sum^{S'\not=
\G}_{S'\in  \ {}_u \!W^h_{1}\{\G\}} %
\left(\Ru t_0<S'>\right)&*\Bigl(\Di(\G)
R\left(W^\tld\{\G,S'\},
\Du\right)<\G/S'>\Bigr)
\EQN{4414}\
\end{align}

The account of the generic properties of the $c$-operation
allows one to replace
the operation $\Di(\G)$ in \re{4414}  by

\beq -\Du(\G)\Rm(W^{\gi}\{\G,S'\},\Du).\nnb \eeq
  
Indeed, according to \re{4411}
\beq \Di(\G)=-\Du(\G)R(F\{\G\},-\Du).\nnb
\eeq
However, if a forest $F\in F\{\G\}$ does not meet the condition
$F\gi S'$, then the expression 
\[\Du(F)\Du(S'')\!<\G/S'>\]
 vanishes for any spinney
$S''$. Thus, in the case under consideration, we may use
\[
\Rm(W^{\gi}\{\G,S'\},\D_U)=R(F^{\gi}\{\G,S'\},-\D_U)
\]
instead of $R(F\{\G\},-\D_U)$. Finally,
on performing the substitution,
there appears a possibility to use \re{gf9} and represent
$R^*t_0<\G>$ in a partially summed  form
\begin{align}
&R^*t_0<\G>=\Ru t_0<\G>\\ 
&+\sum^{S\not=\G}_{S\in {}_u\! W^h_{1}\{\G\}}
\left(\Ru t_0<S>\right)*\bigl(-\Du(\G)R(F^{\succ}\{\G,S\},-\Du)
<\G/S>\bigr)
 \EQN{4415}
\end{align}
or, equivalently,
\beq {R^*t_0<\G>=}\nnb \eeq
\beq \sum_{S\in \   {}_u\!  W^h_{1}\{\G\}}
\left(\Ru t_0<S>\right)*\Bigl(-\D_{\dot
 U}(\G)\sum^{F\not=\G}_{F\in
F^{\succ}\{\G,S\}}(-)^{\vert F \vert}\D_U(F)<\G/S>\Bigr).
\EQN{4416}
\eeq
  

\sect{$R^*$-operation and evaluation of UV and IR
counterterms}

In this subsection we
use the above developed machinery of the $R^*$-operation to
give a simple  proof to an important   theorem
 \ocite{me84} which states that
an arbitrary UV (or IR) counterterm can be expressed in terms
of the  divergent and finite parts of some properly constructed
massless FI's depending on one external momenta.
As a by-product, we also get a new,
practically convenient and shorter version of  the
corresponding algorithm of ref. \ocite{me84}.
\eol
{\bf Theorem  6}. {\it Let \  $t_0\!\!<\G>$ be an arbitrary
dimensionally regularized FI corresponding to a connected
graph $<\G>$ with ${\cal  N}_\G=h$ and with its external momenta
constrained by \re{en1}. Then}

\begin{description}

\item{(a)} The Laurent expansion in $\ep$ of the FI
\ $t_0\!<\G>$ contains only $\ep^i$ with $i\ge -h$;

\item{(b)} {\it Both polynomials
$\ \Du\!\!<G>$ and  $\ \Di\!\!<\G>$ can be
identically expressed via the first
$h$-terms\foot{{\rm By definition, the Laurent
expansion of an $h$-loop  FI starts from the term
$a\ep^{-h}$ even if $a\equiv0$.}}
of the Laurent expansion in
$\ep$  of some massless propagator-type FI's with the
number of loops not exceeding $h-1$.}  \eol
\end{description}
{\bf Proof. }
\begin{description}  
\item{(a)}
This is a well-known  fact in the case of FI's
without IR divergences  which
follows naturally  way from the transition to the
$\alpha$-parametric representation and
a properly chosen change of integration variables (see
e.g. \ocite{BM77a}). Though the argument is no longer
operative in the general case, one  can still  use it to infer
that the polynomial  $\Du\!<\G>$  and, thereby, $\Di\!<\G>$
(as a consequence of \re{4412}) has
poles in $\ep$ no stronger than $\ep^{{\cal  N}(\G)}$ owing to
the  argument presented after Theorem 2 of sect. 3
Thus, if a FI $t_0<\G>$ had a pole at
$\ep\to0$ higher than $\ep^{-{\cal  N}(\G)}$, then
the $R^*$-operation would fail to
 renormalize away {\it all} the poles from
\ $t_0\!<\G>$, which is  in evident   contradiction with
Theorem 5.  
\item{(b)}
The statement is obviously true at $h=1$. Let us prove
it for $h=h_0+1$ assuming
that it has already been proved for all $h\le h_0$.
Owing to \re{4412} it is sufficient to
consider only the case of the UV
counterterm 
\[\Du\!\!<\G>= \sum_i Z_i(\ep){\cal  P}_i({\bf  q},{\bf  m})
.
\]
Without essential loss of
generality, we  may also assume that, first, $\G$ is 1PI
and, second, the FI
$<\G>$ is only log-divergent (that is $\omega(\G)=0$
and $\Du\!<\G>=Z(\ep)$).
Indeed it is well-known \ocite{Vla80a,me80,CK82} that
if $\omega(\G)>0$, then
every renormalization constant $Z_i$ can be expressed
through the UV counterterms of
some set of log-divergent FI's obtained from $<\G>$ by
differentiating
the latter with respect to its external momenta and
masses.
\end{description}  

From Theorem 5 it follows that
\beq Z=-K\left({}'\!\Ru\Rt<\G>\right),\EQN{451}\eeq
or, in explicit form, (see \re{4415})
\beq Z=-K\Bigl[{}'\!\Ru<\G>+\sum^{S\not=\G}_%
{S\in \ {}_u\! W_1\{\G\}}\Ru t_0^{\omega(S)}<S>*%
\Di\!<\G/S>\Bigr]\EQN{451a}\eeq
where we have used the identity
$\omega(\G/S)+\omega(S)=\omega(\G)$
and the definition \re{443}.
$Z$ is a dimensionless polynomial in $\ep^{-1}$,  and
we  hence    have the freedom of setting to zero some (or
even all) external momenta and masses. Now we put
${\bf  q}=0$ and ${\bf  m}=0$ and introduce an non-zero auxiliary mass $\mu_0$
into a (arbitrarily chosen) line $l\in{\cal  L}$
in such a way that the resulting FI $<\G>(\mu_0,\mu,\ep)$ can
be written as\foot{If we did not introduce any auxiliary
momenta or masses after the nullification procedure, the
 rhs of \re{451} would have included the  IR counterterm
$\Di\!\!<\G>$  whose evaluation is as difficult as that of
$\Du\!\!<\G>$.}
\beq <\G>(\mu_0,\mu,\ep)=\int{ <\G'>(k,\mu,\ep)  \over
(k^2+\mu_0^2)}d_\mu k.\EQN{452}
\eeq
Here $<\G'>(k,\mu,\ep)=<\G-l>(k,\mu,\ep)P_l(k)$
with $P_l(q_l)/q_l^2$ being the propagator corresponding
the line $l$ in the initial FI \  $<\G>$.

After the rearrangement of external momenta
and masses has been done, the FI $<\G>$ is to be
naturally interpreted as having ${\bf  q}=\emptyset$ and ${\cal  L}^1=l$.
It is now  clear that every  spinney
from  ${}_u\!W_1^h\{\G\}$ may contain one and only one
graph; the graph must have $l$ among its internal lines
and  must get 1PI after reducing this line.
Moreover, if $\g\in\ {}_u\!W_1^h\{\G\}$ is $not$ 1PI
and $\omega(\g)\ge0$, then the FI
$\Ru t_0^{\omega(\g)}<\g>$ is, in fact, zero.
Indeed, in this case
$\g=\g_1\cup\g_0\cup\g_2$, where $\g_1$ and $\g_2$ are
two disjoint 1PI subgraphs of $\G$
attached to the vertices $\pi_{-}(l)$ and $\pi_{+}(l)$,
respectively, while $\g_0$\   is
the (unique) connected subgraph of $\G$ such that ${\cal  L}(\g_0)=l$.
Since ${\bf  q}=\emptyset$, one has ${\bf  q}_0^{\g}={\bf  q}^{\g}$,
and the only possibility to get a non zero contribution to
$\Ru t_0^{\omega(\g)}<\g>$ is due to the term\foot%
{In general, one of two 1PI graphs $\g_1$ and $\g_2$
might be trivial. In order to take into account
this possibility we shall use  the $\D_{\dot U}$-operation
in the expression below.}
\beq t_0^{\omega(\g)}\left(
(\D_{{}_{\dot U}}<g_1>)<\g_0>(\D_{{}_{\dot U}}<\g_2>)\right).\EQN{new}\eeq
But this expression
is itself equal to zero owing to the fact that
\re{new} is evidently a homogeneous polynomial
in momenta from ${\bf  q}^{\g}$ of degree
$\omega(\g)+2\not=\omega(\g)$!
With  account  of this remark, \re{4415}
 leads to the following simplified
form  of eq.\re{451a}
\beq {Z=-K\Bigl[{}'\!\Ru<\G>\Bigr]
\atop{\displaystyle{-K\Bigl[\sum^{\g\not=\G}%
_{\g}\Ru t_0^{\omega(\g)}<\g>*%
-\Du\!(\G)\sum_{F\in F^\succ\{\G,\g\}}%
(-)^{\vert F\vert}%
\Du\!(F)<\G/\g>\bigr)}\Bigr],}}\EQN{453}\eeq
where the first
sum goes over a 1PI $\g$ such that $\g_0\subset\g$
and $\omega(\g)\ge0$.
On dimensional grounds we have
\beq  <\G'>(k,\mu,\ep)=(\mu/q^2)^{h_0\ep}{P'(k,\ep)\over%
(k^2)^{1+n}},\nnb \eeq
where $P'(k,\ep)$ is a homogeneous polynomial in $k$ of
degree $2n$ with coefficients being some meromorphic
functions of $\ep$. This relation means that the pole part
of the FI\  $<\G>(\mu_0,\mu,\ep)$ itself can be
found by performing a trivial one-loop integration over $k$
in \re{452}, with
the result being expressed through the first $(h_0+1)$ terms
of the Laura expansion of $P'(k,\ep)$ in $\ep$.

In order to finish the proof it remains  to be checked
that the UV c-operation $\Du$  in the rhs of
\re{453}  acts, in fact, only on FI's with the
loop numbers less than or or equal
to $h$. This is, indeed, the case
since  a 1PI graph can not contain   a proper subgraph with the
same number of loops.

Equation \re{453} is a convenient starting point
for evaluating the UV (and thereby IR) counterterms.
Indeed,  the initial
algorithm of ref. \ocite{me84}  relies
on a relation which is nothing but  an  unnecessarily
complicated form of \re{451a}. The advantages of
 \re{453} over the latter are clearly seen. First,
everything here is expressed through the usual UV counterterm
operation; moreover, the calculational procedure
prescribed  by
\re{453} is technically very similar to
the one used in the case where no IR divergences appear.
This means that if someone wants to compute some
UV renormalization constant then   one can hopefully use
eq.\re{453} only
without any need to understand how it is obtained,
not speaking about subtle details of
the $R^*$-operation. Second, the total number
of terms in the  rhs of \re{453} is, in general,
much smaller than in the rhs of \re{451a}.
Finally, eq.\re{453} is applicable without any changes to
calculating UV counterterms in two-dimensional field theories,
while the definitions of ref. \ocite{me84} should be
somewhat modified in this case.

\chap{ ASYMPTOTIC EXPANSIONS OF FEYNMAN INTEGRALS}
This section is mainly devoted to combinatorial  problems
appearing in studying asymptotic behavior of a generic
Feynman integral in the case where
some of its momenta and/or masses go to infinity.

Suppose we are given a decomposition of external momenta and
masses of a dimensionally regularized FI \
$<\G>\!({\bf  q},{\bf  m},\mu,\ep)$, with $\G$ being a connected graph
and ${\bf  q}$ being ${\bf  q}$ constrained by eqs.\re{en1}, namely
\beq {\bf  q}={\bf  q}_2 \cup{\bf  q}_1 \cup{\bf  q}_0 \ ,\  \ \ \ \ \ \ \ \
{\bf  m}={\bf  m}_2 \cup{\bf  m}_1 \cup{\bf  m}_0 ,
\EQN{501}
\eeq
\beq {\bf  q}_i =\{q_l\vert l\in{\cal  E}^i \},\ \ {\bf  m}_i =\{m_l\vert l\in{\cal  L}^i \},\ \
{\cal  E}^i \cap{\cal  E}^j =\emptyset
={\cal  L}^i \cap{\cal  L}^j  \ \hbox{\rm for}\ \ i\not=j ,\ \ i ,j =0 ,1 ,2
,\nnb 
\eeq
where ${\bf  q}_2 ,{\bf  q}_1 $, and ${\bf  q}_0 $ are right subsets of ${\bf  q}$.
Let us denote also 
\beq {\bf  q}_{i\uparrow}=\bigcup_{i'
=i}^{2}{\bf  q}_{i' },\ {\bf  q}_{i\downarrow}=\bigcup_{0}^{i' =i}{\bf  q}_{i'
}, \ {\bf  m}_{i\uparrow}=\bigcup_{i' =i}^{2}{\bf  m}_{i' },\
{\bf  m}_{i\downarrow}=\bigcup_{0}^{i' =i}{\bf  m}_{i' }, \  i=0,1,2\nnb \eeq
  
Note that the sets
${\bf  q}_{i\downarrow}$,
${\bf  q}_{i\uparrow}$,\ \ $i=0,1,2$ are obviously right subsets
of ${\bf  q}$ too. This allows us to   use freely the terminology
introduced in subsection 4.3  with respect to
the partitions 
 
\centerline{${\bf  q}={\bf  q}_i\cup{\bf  q}_{i'},\ {\rm and}\ %
 {\bf  q}={\bf  m}_i\cup{\bf  m}_{i'}$ \ \ \ where \ \ $(i,i')=(2,1\do)$ \ \
or \ \ $(1\up,0)$. \ \ }

\noindent
The problem we are interested in
is to construct an explicitly finite
asymptotic expansion of the  FI
\beq t_{0}\ \llp<\G>({\bf  q},{\bf  m},\mu,\ep,\rho)\equiv
t_{0}<\G>(\lp{\bf  q},\lp{\bf  m},\mu,\ep)\nnb \eeq
as $\rho\rightarrow0$. Here, by definition,
$\lp{\bf  q}=\lp{\bf  q}_2\cup{\bf  q}_1\cup{\bf  q}_0,  \
\lp{\bf  m}=\lp{\bf  m}_2\cup{\bf  m}_1\cup{\bf  m}_0 \ \lp{\bf  q}_2=\{q_l/\rho\vert
q_l\in{\bf  q}_2\}, $
$ {\rm and}\ \ \lp{\bf  m}_2=\{m_l/\rho\vert m_l\in{\bf  m}_2\}$.
Without essential loss of generality
we shall assume that if a vertex $v\in {\cal  V}$ is 2-soft, then
the corresponding vertex polynomial does not depend on the
momenta from the collection ${\bf  q}_2$.
 \sect{Dimensionally
regularized Feynman integrals}
Suppose for a moment that the
FI $<\G>$ does not suffer on any IR divergences. The general
form of the $\rho\rightarrow0$ expansion of the
$R$-normalized\foot {By a $R$-normalized FI we mean the
result of subtracting UV divergences from the FI via an
$R$-operation; the similar convention will be also used for
the $R^*$-operation.} FI  $\ R\,\llp<\G>({\bf  q},{\bf  m})$ considered
in a space-time with an integer dimension
is
 \ocite{ZS65,Za64,Fink68,Sla73,Pohlm82,Berge74,Berge82a,Berge82b}
\beq R\
\llp<\G>\as \ \ \sum^\infty_{i=i_{\rm min}}\rho^{i}
\sum^{{\cal  N}(\G)}_{j=0}(\ln \rho)^j f_{ij}({\bf  q},{\bf  m})
{},
\EQN{511}
\eeq
where $i$ runs over the rational values of an increasing
arithmetic progression.

Now we describe how the expansion \re{511} is naturally
generalized to hold for an unrenormalized FI with all its
divergences being regulated by dimensional regularization.
\endgraf\noindent
{\bf Theorem 7} {\it For arbitrary dimensionally
regularized Feynman integral
$t_0\,'\!\!\!<\G>({\bf  q},{\bf  m},\mu,\ep,\rho)$ \eol
there holds the
asymptotic expansion as $\rho\to0$ of the form
\beq t_0\,'\!\!\!<\G>\as \ \ \sum^\infty_{i=i_{\rm  min}} \ \
\sum^{{\cal  N}(\G)}_{j=0}\rho^{i-2j\ep} \
F_{ij}({\bf  q},{\bf  m},\mu,\ep)\EQN{512}\eeq
   where the functions
$F_{ij}$'s are meromorphic in $\ep$ and homogeneous with
respect to the momenta and masses from the collections
${\bf  q}_2$ and ${\bf  m}_2$ respectively, to wit:
\beq F_{ij}(\lp{\bf  q},\lp{\bf  m},\mu,\ep)=\rho^{i-2j\ep}F_{ij}
({\bf  q},{\bf  m},\mu,\ep)\EQN{513}\eeq
   \endgraf\noindent
The expansion \re{512} remains valid
after its left and right parts are
both subjected to the Laurent expansion in $\ep$.}

The proof of the theorem generalizes and to some extent
repeats the reasoning given in
refs. \ocite{Berge74,Pohlm82,BM77a,BM77b,BM77c}. Since
it is rather lengthy and not especially instructive, it will
be given elsewhere.

Our next task is to construct an explicit representation of
the rhs of \re{512} in terms of subgraphs of $<\G>$ and
the respective reduced graphs. Let us define the {\it
glued}\foot {The term comes from
works \ocite{me82b,me88a,me88b} that deal with a
particular kinematical regime where only one independent
external momentum goes to $\infty$.} FI $\
t_0\!<\hat\G>({\bf  q},{\bf  m},\ep,\mu,\delta)$ as a Mellin transform of
$\theta(\rho-1)t_0\,\llp<\G>({\bf  q},m,\mu,\ep,\rho)$ with
respect to $\rho$, that is
\beq t_0<\hat\G>=\int_0^1t_0\llp<\G>\rho^{-\delta-1}d\rho\EQN{514}
\eeq
Due to Theorem 7 the integral $<\hat\G>$ is
a meromorphic function of $\delta$ with simple poles located at
$\delta=\delta_{i,j}\equiv i-2j\ep$; the  respective residues are
proportional to the functions $F_{ij}$.

On the other hand, within the $\alpha$ representation technique
the glued FI \re{514} is akin to usual dimensionally
{\it and} analytically regularized FI provided the parameter
$\rho$ is treated as an extra $\alpha$-parameter.
This key observation allows one to employ the existing
powerful methods of studying  analytical structure of such
integrals \ocite{Sp69,BM77a,Smi85,Smi91}.  The result is
described by the following\foot {An elaborated proof of the
theorem has been found within the outlined approach by the
present author and will be published elsewhere.}.
\vspace{3mm}

\noindent
{\bf Theorem 8.}  {\it The expansion
\re{512} can  be identically rearranged so that it takes
the form
\beq t_0 \ \llp<\G>\as\sum_{ \ \ \ S\in \  {}_u\!
W_2\{\G\}} t_{1\do} \ \llp<S>*<\G/S>,\EQN{515}\eeq
  
where \ \
$\displaystyle\llp<S>\equiv\prod_{\g\in S} \  \ \llp<\g>$,
with \endgraf\noindent \centerline{
$\llp<\g>\equiv<\g>(\lp{\bf  q}^\g, \ \lp{\bf  m}^\g)$ \ \ \hbox{and} \
\ $\lp{\bf  q}^\g\equiv({\bf  q}_2/\rho)\cup{\bf  q}^\g_{1\do}$, \ \
$\lp{\bf  m}^\g\equiv ({\bf  m}^\g_2/\rho)\cup{\bf  m}^\g_{1\do}$.}}

\noindent 
There are several remarks we would like to make in
connection with the theorem.  
\begin{description}
\item{(i)} All the dependence
on $\rho$ in the rhs of \re{515} is located in the first
factor, with
\beq t_{1\do}^{(n)} \,
\llp<\g>=\rho^{-\omega(\g)+2\ep{\cal  N}(\g)+n}<\g>
{}.
\eeq
   
\item{(ii)} Theorem 7  ensures that the
expansion \re{515} does commute with expanding in $\ep$.
This fact will be of vital importance below in proving
Theorem 5.
\item{(iii)} The sum in \re{515} would not change if it
had gone over $S\in {}_u W^h_2\{\G\}$. Indeed, if $S$
contains a 2-soft element $\g$, then the factor
$t_{\do}<\g>$ vanishes since it is a linear combination of
the massless tadpoles.
\end{description}
Finally, as a direct consequence of the above theorem we find that
the following statement holds.
\vspace{1mm}

\noindent
{\bf  Theorem 9}
\begin{description}
{\it \item {(a)} If $\g$ is a  $2$-right  subgraph
of $\G$,  then the $\rho\to0$ asymptotic expansion of the
respective Feynman subintegral can be written as
\beq t_0 \ \llp<\g>\as\sum_{ \ \ \ S\in \>  {}_u\! W_2\{\g\}}
t_{1\do} \ \llp<S>*\>t_0\!<\g/S>.\EQN{516}\eeq

\item{(b)}
If $\Phi$ is an arbitrary spinney from
$\Wu\{\G\}$, then the asymptotic expansion of
the FI $<\G/\Phi>$ as $\rho\to0$  assumes the form
\beq t_0 \ \llp<\G/S>\as\sum_{ \ \ \
S\in \  {}_u\! W_2\{\G\}}^ {S{\ge}{\Phi}}
t_{1\do} \> \llp<S/\Phi>*\>t_0\!<\G/S>.\EQN{517}\eeq

}  
\end{description}
\noindent
{ {\bf Proof. }}
\begin{description}

\item{(a)}  Obvious.

\item{(b)} There exists a one-to-one
correspondence between two collections of spinneys
\[\{S\vert S{\ge}\Phi,S\in {}_u\! W_2\{\G\}\]
and
\[ \{S'\vert S'\in {}_u\! W_2\{\G/S\},\]
 which is established by
the mapping (cf. the proof of \re{r4})
\beq r:\ S {}^{\phantom{aaaa}r}{\!\!\!\!\!\!\!\longrightarrow} \ S'=
(S\backslash\Phi)/\Phi.\EQN{518}\eeq

Relation \re{517} follows from this correspondence and
the observation that
\beq <(S\b\Phi)/\Phi>=<S/\Phi>\nnb
{}.
 \eeq

\end{description}

\sect{$R^*$-normalized Feynman integrals}

We now turn to to finding the $\rho \to0$ asymptotic
expansion of the
$R^*$-normalized FI
\beq {R^*\>t_0\llp<\G>=
R^*(\G) \ t_0\,<\G>(\lp{\bf  q},\lp{\bf  m},\mu,\ep)=\atop{{}\atop
\displaystyle{\sum_{\phantom{a}\Phi\in {}_u\! W_{1\up}\{\G\}}
\Ru t_0\,\llp<\Phi>*\D_{{}_{\dot I}}<\G/\Phi>.}}}\EQN{521}\eeq
This is certainly a purely algebraic problem since the FI
\re{521} is virtually nothing but a linear combination of
unrenormalized FI's multiplied by some UV and IR renormalization
constants and  its solution will hence  rely upon
the forest technique developed above.

To begin, we assume for a while that
${\bf  m}_0={\bf  q}_0=\emptyset$ and, thus, the operator $t_0$ in the first
line of \re{521} reduces to the unit one.
This means in particular that the FI
$<\G>$ does not contain any IR divergences
\foot{Recall that as in our conventions
{\it all} massless lines are supposed to be assigned to
${\cal  L}^0$.}
and the $R^*$-operation may be safely replaced by the $R$-one.
Under the circumstances the asymptotic expansion we are looking for
can be directly read off from \re{517}:
\beq {\displaystyle{{R<\G>\as \  \
\sum_{\ \ S_a\in \  {}_u\! W_2^h\{\G\}}
\ \ \sum_{S_b\in W_2^s\{\G\}}^{S_b\sim S_a}
\ \ \sum_{\phantom{a}\Phi_a\in W_{UV}\{\G\}}^{\Phi_a\le S_a}
\ \ \sum_{\phantom{a}\Phi_b\in W_{UV}\{\G\}}^{\Phi_b\le S_b}}}
\atop{{}\atop\displaystyle
\Du<\Phi_a>*\>t_{1\do}\,\llp<S_a/\Phi_a>
*\Du<\Phi_b>*t_{1\do}\,<S_b/\Phi_b>*<\G/(S_a\cup S_b>,}}
\EQN{522}\eeq
  
As explained after Theorem 8
the  terms with $S_b\not=\Phi_b$ do  not  contribute
to \re{522}. This allows us
to rewrite \re{522} in the following compact form
\beq R<\G>\as\sum_{\phantom{a}S \in\  {}_u\! W_2^h\{\G\}}R\,t_{1\do}\llp<S>*
R(W^\sim\{\G,S\},\Du)<\G/S>.
\EQN{523}\eeq

The  asymptotic expansion is {\it not} explicitly finite:
both factors $R\,t_{1\do}\llp<S>$
and $R(W^\sim\{\G,S\},\Du)<\G/S>$ do in general
suffer from the IR and UV divergences respectively.
On the other hand these divergences should cancel out
owing to the finiteness of the initial (renormalized!) FI
$R<\G>$ at $\ep\to0$. Hence, there should exist an explicitly
finite version of \re{523}. The problem can be solved
by simple adding the missing subtractions to the both
factors in the rhs of \re{523} followed by a proof that
the transformation is in fact an identical
rearrangement of terms of the expansion.


We begin with  two identities (see eqs.(\ref{gf7} - \ref{gf9})
\begin{subequations}\label{5.24}
\beq R_U(W^\sim\{\G,S\})=R_U(W^{\gi} \{\G,S\})
R^{-1}_U(W^\succ\{\G,S\}) \EQN{524a}\eeq
and
\beq R^{-1}_U(W^\succ\{\G,S\})=R(W^\succ(\G,S),\D'_I),\EQN{524b}\eeq
where we have denoted
\beq \D'_I(\g)=-\Du(\g)\sum_{\phantom{a}F\in F^\succ\{\G,S\}}
^{F<\g}
(-)^{\vert F \vert}
\D_U(F),\EQN{524c}
\eeq
\end{subequations} 
\noindent
and $R_U(W^\sim\{\G,S\}) \equiv R(W^\sim\{\G,S\},\Du)$ and so on.
Note also that due to Theorem 1 for any spinney $S'\succ S$ one has
\beq \Ru(W^{\gi}\{\G,S\})<\G/(S'\cup S)_{\max}>=
\Ru<\G/(S'\cup S)_{\max}>.\EQN{525}
\eeq
  
We now use (\ref{5.24},\ref{525}) in order to transform the rhs of
\re{523} into
\beq \sum_{S \in\  {}_u\! W_2^h\{\G\}}
\sum_{\phantom{a}S' \in\   W\{\G\}}^{S'\succ S}
\Ru\,t_{1\do}\llp<S>*\Di'<S'/S>*
\Ru<\G/S''>,\EQN{526}\eeq
where $S''=(S'\cup S)_{\max}$.
If $S'$ includes a 2-reducible graph $\g'$, then
the corresponding term in \re{526} can be safely dropped due
to appearance of the vanishing factor
$\D_{{}_{I}}'<\g'/S>$. Hence only spinneys
$S'\in W_2^h\{\G\}$ do contribute to \re{526}.
Next, after going to the summation over $S''$
and employing
the identity
\[\Di'<S'/S>=\D_{{}_{\dot I}}'<S''/S>\]
we have
\beq \Ru<\G>\as \ \sum_{S,S'' \in\  {}_u\! W_2^h\{\G\}}^{S\le S''}
\Ru\,t_{1\do}\llp<S>*\D_{{}_{\dot I}}'<S''/S>*
\Ru<\G/S''>\EQN{527}\eeq
or, due to the relation \re{4416},
\beq \Ru<\G>\as \ \sum_{S'' \in\  {}_u\! W_2^h\{\G\}}
R^*\,t_{1\do}\llp<S''>*
\Ru<\G/S''>,\EQN{528}\eeq
which is the desired {\it explicitly finite} writing
of \re{523}.

In order to cover the general case of the expansion of the
$R^*$-normalized FI  \re{521} one should learn to expand
products like
\beq \Ru t_0\,\llp<\Phi>=\prod_{\g\in \Phi}\Ru t_0\,\llp<\g>,\EQN{529}\eeq
with $\Phi\in \ {}_u W_1\{\G\}$. Fortunately,
the result of expanding in $\rho$ every factor
in the rhs of \re{529} may be directly obtained
with the help of \re{528} since its derivation has
used neither IR finiteness of $<\G>$
nor the absence of the operator $t_0$. As  a result, we get
\beq \displaystyle{R^*t_0\,\llp<\G>\as \
\sum_{\Phi\in{}_u\!W_{1\up}\{\G\}}
\sum_{\phantom{aaa}S \in\  {}_u\! W_2^h\{\G\}}^{S\le\Phi}}
\atop{{}\atop\dsp
{R^*\,t_{1\do}\,\llp<S>*\Ru t_0\,<\Phi/S>*
\D_{{}_{\dot I}}<\G/\Phi>,}}\EQN{5210}\eeq
which, in turn, can be considered as
an alternative  form of the
following remarkable asymptotic expansion
\beq R^*t_0\,\llp<\G>\as \ \sum_{S \in\  {}_u\! W_2^h\{\G\}}
R^*\,t_{1\do}\llp<S>*
R^*t_0\,<\G/S>.\EQN{5211}\eeq
Indeed, to derive \re{5211}
from \re{5210} it suffices to observe, first,  the existence of
a natural one-to-one correspondence between
two woods 
\[\{\Phi\vert\Phi\in {}_u W_{1\up}\{\G\},\Phi\ge S\}
\text { \ and \ }
\{\Phi'\vert\Phi\in {}_u W_{1\up}\{\G/S\}\}
\]
with $\Phi'=(\Phi\b S)/S$ and, second,
to note that
$<(\Phi\b S)/S>=<\Phi/S>$.

We conclude  this subsection by  presenting a simple proof of
the fact that the $R^*$-operation does subtracts all
kinds of divergences from a generic FI $t_{1\do}<\G>$
(Theorem 5).

Let us introduce an auxiliary positive mass $\mu_0$ into
all the propagators from ${\cal  L}^{1\do}$ and assume that $\mu_0$
is {\it not} subjected to the Taylor expansion under
the action of the operator $t_{1\do}$. Evidently, the
presence of the auxiliary mass prevents any IR divergences
from appearing and thus FI $\Ru t_{1\do}^{(n)}\llp<G>$
is finite at $\ep\to0$ for every integer $n\ge0$. It follows
from \re{5211} that
\beq \Ru t_{1\do}^{(n)}\,\llp<G>\as\
\sum^{k+k'\le K}_{k,k'\ge0}\sum_{\phantom{aa}S\in
{}_u W^h_2\{\G\}}
^{S\not=\G}R^*\,t_{1\do}^{(k)}\T_{\mu_0}^{(k')}\ \llp
<S>*\Ru t_{1\do}^n<\G/S>\nnb \eeq
\beq + \sum^{k+k'\le K}_{k,k'\ge0}R^*t_{1\do}^{(k)}
\T_{\mu_0}^{(k')}\,\llp<\G>+
\delta(\rho), \EQN{531}
\eeq
where, due to Theorems 7 and 8, 
the reminder\foot{Here $\displaystyle{N=\max_{S\in{}_u\!W^h_2\{\G\}}
\left(\sum_{\g\in S}\omega (\g)\right)}$
and  $\ep$ is understood to be sufficiently small and negative.}
$\delta(\rho)=o(\rho^{-N+K+2{\cal  N}(\G)\ep})$
with
at $\rho\to0$
is an analytical function of
$\ep$ in the vicinity of the point $\ep=0$.
Now, reasoning by recurrence with reference to
the number of internal lines of $\G$ one can easily
convince oneself that the sum
\[{\sum^{k+k'\le K}_{k,k'\ge0}R^*t_{1\do}^{(k)}
\T_{\mu_0}^{(k')}\,\llp<\G>}
\]
is finite as $\ep\to0$, because of the finiteness of the left
side of \re{531}.
This, in turn, means that every
term in this sum is finite for the values of  $\rho$ and
$\mu_0$ can be chosen at will. Finally, choosing
$\rho=1,k=n$ and $k'=0$  we conclude that the FI
$R^*\Bigl[t_{1\do}^{(n)}<\G>\vert_{\mu_0=0}\Bigr]$
is also finite as $\ep\to0$, which was the thing to be proved.
\chap{RELATED WORKS: A COMPARATIVE \\ DISCUSSION OF
 RESULTS}
{\it $\Rm$-operation.}
There exists a deep similarity between the $\Rm$-operation
technique and the counterterm formalism \ocite{AZ76,AZP77,An82,Za89}. 
To some
extent,  the former may be considered as another variant
of the latter adopted to be able to deals with
with separate FI's with the same ease as the counterterm
formalism treats  the perturbation series as a  whole.

{\it Infrared rearrangement and $R^*$-operation}.
After the pioneering work by
Vla\-di\-mi\-rov \ocite{Vla80a}, the trick of infrared
rearrangement has been  repeatedly rediscovering by many
authors (see e.g. refs. \ocite{Damme84,Marcus85}). The authors of
the second work even claim that they have proved a statement
analogous to Theorem but their proof makes no allowance
for the IR divergences.  Certainly in the absence
of any IR subtractions such a statement may be true
provided that the FI obtained after the IRR has no IR
divergences.  This, in turn, means that the procedure
suggested in ref. \ocite{Marcus85} is completely equivalent to
the IRR trick as it was described in ref. \ocite{Vla80a}.

A subtraction procedure  to remove  the IR divergences which
appear if one tries to expand a convergent FI with
non-exceptional external momenta when all its masses go to
zero, has been discussed in ref. \ocite{Be82}. The
procedure is, in fact, a very particular version of the
$R^*$-operation:  the assumed UV finiteness of the integral
to be expanded makes things  much easier and,
simultaneously, restricts severely the possible
applications.  Indeed, both of the most useful applications
of the $R^*$-operation --- calculation of the
renormalization constants and construction of  explicitly
finite asymptotic expansions of generic FI's --- seem to be
lost.

The natural normalization condition \re{449} for the
$R^*$-operation  was first suggested in \ocite{me83b} with
essentially the same motivation as the one put forward in
 the  present work.

The  accurate proof of Theorem 5 --- the main statement
of the $R^*$-operation   theory --- has been
presented in \ocite{me89a} for the  particular case of a FI
with nonexceptional external momenta. The proof is based on
the heavy use of the $\alpha$-parametric representation and
 technical tools developed early
in refs. \ocite{Sp69,BM77a,Smi85} ( for an up-to-date review see
a book \ocite{Smi91}).

{\it Asymptotic expansions}.
There are a few series of studies devoted to the
problem of constructing a {\it complete} asymptotic
expansion of a generic dimensionally regularized euclidean
FI when some of its external momenta and/or masses go to
infinity. The authors of
refs. \ocite{T84a,T84b,T84c,T86,T88}
have investigated the problem within a so-called "extension
principle" (EP) which, in fact, was inspired by the ideas of
the $R^*$-operation as formulated in ref. \ocite{me82a}. In
particular, in ref. \ocite{T84c} an expansion equivalent to
the one \re{523} has been suggested as a natural generalization of
the results obtained by applying the EP to a few particular
one-loop FI's. As far as we  know, no proof of this
expansion or of the EP itself has been published yet.
Moreover it has been explicitly stated in \ocite{T84c} that
 the EP is {\it not} applicable to FI's with IR
divergences. This means that the approach fails to expand a
generic $R^*$-normalized FI.

In his two  works \ocite{Gor86}, the late Doctor S.G. 
 Gorishny suggested a method of constructing asymptotic
expansions of FI's at large momenta and/or masses. His
method is based on a heuristic generalization of the
Zimmermann approach \ocite{Z70,Z73b}. Its starting point
is an asymptotic expansion analogous to  \re{523} in which,
however, the  operator  $t_0$ expands in momenta from
${\bf  q}_0$ around some fixed {\it non-zero} point in momentum
space. This modification is introduced to avoid the IR
divergences.  After developing a combinatorial technique
similar to ours, the author of ref. \ocite{Gor86} found an
expansion which should presumably be equivalent (modulo the
different definition of the $t_0$ - operator) to eq.\re{528}.
He also gave the heuristic argument that within the MS -
scheme the choice of the zero momentum expansion point
should not lead to IR divergences. Note also work \ocite{LLS88}
similar in its spirit and results,  which, however,
deals with a very  particular case of the problem under
discussion, namely with the  short distance expansion of a
product of two composite operators.

The fact that it is the $R^*$-operation which naturally
appears when one treats asymptotic expansions of
minimally renormalized FI's was  first realized
in refs. \ocite{T83a,me83b}. The expansions
\re{523} and \re{528} were
derived for the case where   only one external
momentum is considered to be  large in 
refs. \ocite{me83b,me88a,me88b}. Here the Wilson short
distance expansion of a product of two composite operators
was studied in the framework of the gluing
method \ocite{me82b}.

A heuristic argument in favor of the existence of the expansion
\re{528} was suggested in ref. \ocite{GorLar87}.

A practically convenient algorithm to compute
the coefficient functions  of various operator expansions
in the MS-scheme was presented in refs. \ocite{T83b,GorLar87}.
At the level of individual Feynman graphs the algorithm amounts, in fact,
to the expansion \re{523}, which is taken as granted.

Recently a rigorous proof of the expansion \re{528} has
been obtained in  \ocite{Smi88,Smi89} along the lines of
refs. \ocite{Z70,Z73b,AZ76b,Za89}, by constructing a
suitable oversubtraction operator which makes  use of the
$\Rt$-operation.
In addition, in these works     the counterterm
formalism was effectively employed
to derive explicitly finite asymptotic expansions of
minimally renormalized Green functions for a variety of
physically interesting asymptotic regimes.
Note that the generalization of the
approach to generic $R^*$-normalized FI's seems to be
nontrivial (see in this connection ref. \ocite{me87a}).
\chap{CONCLUSIONS}
In this work we have elaborated a technique which has
allowed us to perform a uniform, self contained and
essentially complete study of a variety combinatorial issues
involved in the $R-,\ R^{-1}-\  {\rm and}\  R^*$-operations and their
applications. We find it nice that all the three
 "renormalization" operations  are tightly interconnected
each other, with the interplay showing up not only in
the MS - scheme (what is more-or-less natural, since the
inverted and the starred R-operations first appeared in the
context of just this scheme) but also in a different
renormalization prescription --- the momentum subtraction
scheme.

There remain few interesting problems that can
be solved within the approach. Below we list
some of them.
\endgraf
{\it $\Rm$-operation}. To
construct practically convenient conversion  formulas which
are to connect the standard MS - scheme with its close
relative based on so-called dimensional reduction.
 By finding a suitable
conversion formula
to clarify the relationship between different definitions of
the $\g_5$-matrix in DR .

\endgraf
{\it$\Rm$- and  $R^*$-operations.} It would be
interesting to find a generalization of the relations
\re{3.39} between momentum subtractions
and the $\Rm$-operation
in the case of renormalization with the soft mass.
In doing the problem a kind of $R^*$-operation
should presumably  appear in a natural way.

The $R^*$-operation is a natural tool to investigate the
IR finiteness of the given theory including massless
particles. With its help it is possible e.g. to give a
 simple and purely combinatorial proof of the  IR finiteness
of quantum chromodynamics along the way outlined
in refs. \ocite{BM77a,BM77b,BM77c}.
This has been done by the
present author and will be published elsewhere.
\endgraf {\it Asymptotic expansions.} It would be of
interest to construct an explicit and purely combinatorial
proof the renormalization group equation to which the $R^*$ -
normalized  FI's should satisfy.

{\it Acknowledgments.}  I would like to thank Profs.
P. Breitenlohner, D. Maison, K. Zibold and W. Zimmermann
for their interest and discussions. I deeply appreciate the warm
hospitality of all the members of the Theoretical group
of the Werner-Heisenberg-Institut f\" ur Physik where
the substantial part of this work has been made.
I am grateful to V.A. Smirnov for his continuous interest and
encouragement.

\newcommand{\subs}{\textup{s}}


\section{Comments}

The work has never been submitted for publication for purely personal reasons.
Now, a quarter of century later, I want to make it more accessible 
as it is the improved and extended formulation of the
$R^*$-operation presented here has been one of the crucial tool 
(along with the algebraic manipulation language FORM 
[83,84] and other advances in our understanding of multiloop Feynman diagrams (see, e.g. [85])
for many record-breaking multiloop calculations in gauge theories performed during last years 
[86-92].

I want to thank Franz Herzog, Misha Kompaniets, Ben Ruijl, Takahiro Ueda, 
Jos Vermaseren and  Andreas Vogt for their interest to the $R^*$-operation which has stimulated
me to prepare this publication. 

I am grateful to Misha Kompaniets for   attentive reading the text
and sending me his comments.

\ed